\documentclass[aps,prb]{revtex4}
\usepackage{amssymb,bm}
\usepackage{amsmath}
\usepackage[mathscr]{eucal}

\usepackage{graphicx} 

\begin{document}
\title{Nonlinear Sigma Model for Normal and Superconducting Systems: A Pedestrian
Approach}

\author{Igor V. Lerner}

\affiliation{School of Physics and Astronomy, University of
Birmingham, B15 2TT, UK}

%

\def\ee{{\mathrm e}}
\def\dd{{\mathrm d}}
\def\ub#1#2{{\underset{{#1}}{\underbrace{{\, #2\, }}}}}
\def\D {{\mathcal D}}
\def\G {{\mathcal G}}
\def\cS {{\mathcal S}}
\def\cF {{\mathcal F}}
\def\Sp{\textsf{Sp}\,}
\newcommand\U{\textsf{U}\,}
\newcommand\SO{\textsf{SO}\,}
\newcommand\SU{\textsf{SU}\,}
\newcommand\psiint{\int\!\! \D \psi^* \D \psi\,\,}
\newcommand\Psiint{\int\!\! \D {\overline \psi} \D \psi\,\,}
\newcommand\bpsiint{\int\!\! \D {\overline {\bm\psi}} \D \bm\psi\,\,}
\newcommand\bPsiint{\int\!\! \D {\overline { \Psi}} \D  \Psi\,\,}
\newcommand\opsi {{\overline {\psi}}}
\newcommand\opsip {{\overline {\psi}}^{\,+}}
\newcommand\opsim {{\overline {\psi}}^{\,-}}
\newcommand\Psib {{\overline { \Psi}}}
\newcommand\bPsi {{{ \Psi}}}
\def\S{\textsf{S}\,}
\newcommand\diag{\operatorname{diag}}
\newcommand\Tr{\operatorname{Tr}}
\newcommand\tr{\operatorname{tr}}
\newcommand\ef{{\varepsilon_{{\text{{ F}}}}}}
\newcommand\pf{{p_{\,{\text{{F}\!}}}}}
\newcommand\vf{\upsilon_{{\text{{F}\!}}}}
\def\Eq#1{{Eq.\!~(\ref{#1})}}
\def\Ref#1{(\ref{#1})}
\def\Cite#1{ Ref.~\cite{#1}}
\def\Eqs#1{{Eqs.\!~(\ref{#1})}}
\def\EQS#1#2{{Eqs.\!~(\ref{#1}) and (\ref{#2})}}
\newcommand\EQ[2]{{Eqs.\!~(\ref{#1})--(\ref{#2})}}
\newcommand\E[1]{\!#1\!}
\newcommand\Fra[5]{\left #1 \frac {#3}{#4} \right #2^{\!#5}}
\def\lr#1#2#3{\left#1{#3}\right#2}
\def\Bg#1#2#3{\Big#1{#3}\Big#2}
\newcommand\Int{\int\!\!}
\newcommand\spr[2]{{\mathbf {#1}} \cdot {\mathbf {#2}}}
\def\av#1{\left <{#1}\right >}
\def\b#1{{\bm {#1}}}
\def\bra#1{\left\langle {#1}\right|}
\def\Lr#1#2#3#4{\left#1{#3}\right#2^{\!#4}}
\def\ket#1{\left| {#1}\right\rangle}
\def\tel{\tau_{\text{el}}}
\def\braket#1#2{\left\langle {#1}\,|\, {#2}\right\rangle}
 \newcommand\ir[1]{\int\!\!{#1}{\mathrm d}^dr}
 \newcommand\ip[1]{\int\!\!{#1}\frac{{\mathrm d}^dp}{(2\pi)^d}}
\def\dff#1#2#3{\frac{\delta ^2  #1}{\delta {#2}\, \delta {#3}}}
\def\diff#1#2{\frac{\partial^2 }{\partial {#1}\, \partial {#2}}}
\newcommand\aB{{a_{ {\text{{ B\!}}}}}}
\newcommand\lf{{\lambda_{\text F\!}}}
\newcommand\const{\text{const}}
\newcommand\Ud{U^\dagger}
\newcommand\br{{\bm r}}
\def\Ascr{{\ensuremath{\mathscr  A}}}



\def\text#1{{\mathrm {#1}}}
\def\bbox#1{{\mathbf {#1}}}
\def\N{NL$\sigma$M}
\def\SM{$\sigma$ model}
\def\openone{\leavevmode\hbox{\small1\kern-3.3pt\normalsize1}}%

\begin{abstract}
The nonlinear $\sigma$ model (NL$\sigma$M) epitomises a
field-theoretical approach to (interacting) electrons in
disordered media. These lectures are aimed at the audience who
might have vaguely heard about the existence of the NL$\sigma$M
but know very little of {\it what} is that, even less so of {\it
why}\ it should be used and next to nothing of {\it how} it can be
applied. These {\it what, why} and mainly {\it how} are the
subject of the present lectures. In the first part, after a short
description of {\it why} to be bothered, the \N\ is derived from
scratch in a relatively simple (but still rather mathematical) way
for non-interacting electrons in the presence of disorder, and
some illustration of its perturbative usage is given.  In the
second part it is generalised, not without some leap of faith, to
include the Coulomb repulsion and superconducting pairing.
\end{abstract}

\maketitle

\section{Introduction}

Starting from the seminal paper by Wegner \cite{Weg:79}, a
field-theoretical description based on the nonlinear $\sigma$
model (NL$\sigma$M) has become one of the main analytical
approaches to quantum transport and thermodynamics in disordered
electronic systems [2--24].
 Quantum effects are due to
interference of electron scattering from different impurities.
These effects are relevant when the scattering mean free path
$\ell$ is small compared to the dephasing length $L_\phi$ which
characterises the scale above which interference is destroyed by
an irreversible phase-breaking in electron's wave function due to
inelastic interactions. As $L_\phi$ diverges when $T\!\to\!0$, the
quantum interference effects are always important at low
temperatures. Although the quantum corrections to transport
coefficients are small in metals as a power of the small ratio
$\lf/\ell$ ($\lf$ is the Fermi wavelength), they are essential in
electron transport as they govern a nontrivial dependence on
temperature, frequency or external fields \cite{AALR,GLKh}.
Furthermore, the electron-electron interaction in the presence of
the scattering from impurities also results in similar (small in
$\lf/\ell$) corrections both to the transport coefficients and to
thermodynamic quantities (see Refs.~\cite{AA1,LR} for reviews), in
a stark contrast to the case of pure normal metals where the
interaction manifests itself only via the Fermi liquid parameters,
which renormalise electrons spectral coefficients. The \N\ gives
analytical tools for describing both the weak disorder regime
($\lf\ll\ell$) and a regime when the disorder increases with
$\lf/\ell$ approaching a critical value of order $1$ where
electronic states at the Fermi energy are localised
\cite{And:58,IR:60}.

There are two main characteristic features in the $\sigma$-model
approach to the  disordered system: (i) one performs the ensemble
averaging over all configurations of disorder at the first stage
of calculations; (ii) one separates ``fast'' and ``slow" degrees
of freedom and deals only with the latter. In the present context,
``fast'' implies energies of order $\ef$ and momenta of order
$\pf$ while ``slow" means energies smaller than $\hbar/\tel\equiv
\hbar\vf/\ell$ and momenta smaller than $\hbar/\ell$. Such a scale
separation works for a weak disorder. Quite often, the weak
disorder condition $\lf\ll\ell$ is formulated as $g_0\gg1$ where
$g$ is the dimensionless conductance (loosely, the conductance
measured in units of $e^2/\hbar$) and $g_0\propto (\ef\tel)^{d-1}$
is the dimensionless conductance of a cube of size $\ell^d$ in a
$d$-dimensional system. The slow modes in a system of size
$L\gg\ell$ describe diffusive propagation of electrons, and the
standard \N\ is the theory of interacting diffusive modes
(diffusons).\cite{note}

The functional of the $\sigma$ model has the following form:
\begin{equation}
\label{nlsm} {\cal F}[Q;\omega]=\int\!\! \text{d}^dr\,
\text{tr}\left[\frac{\pi \nu_0 D}8(\nabla
Q)^2-\frac{i\pi\nu_0\omega }4\Lambda Q\right]\,,
\end{equation}
where $D\!=\!\vf^2\tel/d$ is the diffusion coefficient, $\nu_0$ is
the one-electron density of states, $d$ is the spatial
dimensionality and $\omega$ is the frequency. The Hermitian matrix
field $Q(\bm r)$ (whose symmetry and dimensions depend on this or
that particular way of the ensemble averaging) obey the following
constraints which make the model nonlinear:
\begin{equation}\label{constraint}
    Q^2=\openone\,,\qquad \Tr Q=0\,.
\end{equation}
Writing down the above functional hardly explains for
non-initiated what is the meaning of the model. It allows me,
however, to formulate the main goals of these
lectures:\vspace{-2mm}

\begin{enumerate}\addtolength\itemsep {-1mm}
  \item To outline a pedestrian but consistent step-by-step
  derivation of the
  above functional for non-interacting electrons in a
  random potential.

  \item To generalise the model, with the help of one or two
  leaps of faith, for including the Coulomb repulsion and
  pairing attraction of electrons.

  \item To illustrate using one or two relatively simple examples
   how the NL$\sigma$M works, both for noninteracting and
   interacting electrons.
\end{enumerate}

\noindent But before attending to this programme,  it is useful to
give some ideas of {\it why} the \N is to be used at all, or where
there are its advantages as compared to the usage of other
analytical means, like a straightforward diagrammatic approach or
a semi-phenomenological random matrix theory (whose applicability
to the description of non-interacting electrons in disordered
systems in a so-called ergodic regime has been rigourously
established with the help of the \N\ \cite{Ef:82,VWZ}).

\begin{figure}[b]
   \centerline{\includegraphics[width=80mm,angle=0]{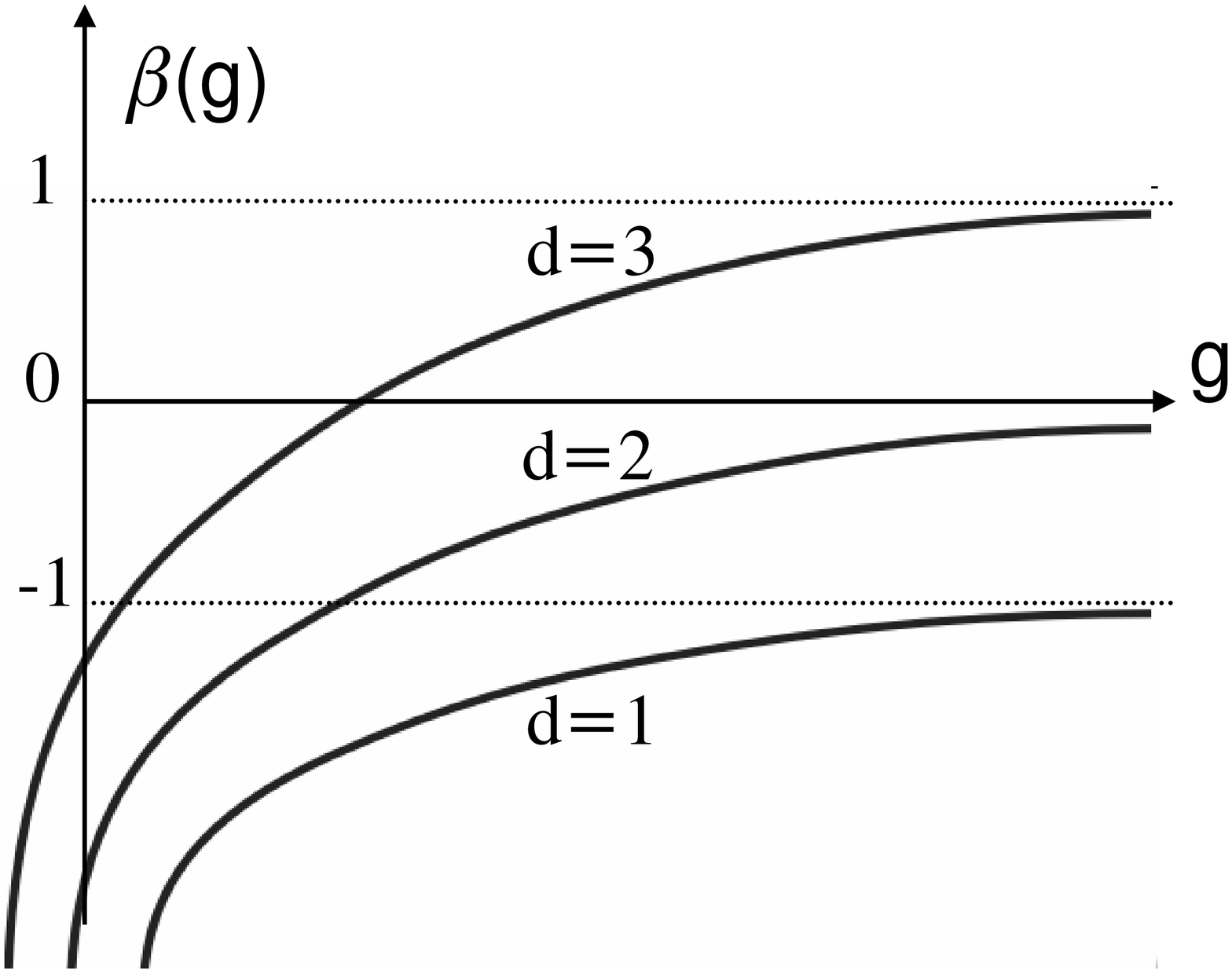}}
   \caption{Schematic shape of the
   $\beta$ function in the one-parameter scaling theory}
\label{beta} \end{figure}

The first application of the \N\ was a derivation 
\cite{Sch+W,MK+S,EfLKh,Ef:82} of the renormal\-ization-group (RG)
equations of the one-parameter scaling theory \cite{AALR} of the
Anderson localisation, illustrated on Fig.~\ref{beta}.
The $\beta$ function, $\beta(g)\equiv \dd \ln g/\dd \ln (L/\ell)$,
depends only on the dimensionless conductance $g$ and describes
how it changes with scale or disorder. For $d=3$, conductance
increases with the system size for $\beta(g)>0$ and decreases for
$\beta(g)<0$, becoming an ideal metal or ideal insulator,
respectively, in the limit $L\to\infty$. In this case $\beta(g)=0$
corresponds to the critical point for the metal-insulator
transition. For $d=2$ and $d=1$, $\beta(g)$ is always negative so
that no truly metallic state can exist in the thermodynamic limit.
However, scaling considerations alone could only prove \cite{AALR}
that $\beta(g)=(d-2)g$ for $g\gg1$, thus leaving some uncertainty
in the marginal $d=2$ case. The uncertainty has almost immediately
been resolved \cite{GLKh} with the help of the diagrammatic
perturbation theory by proving that $g/g_0=1-g_0^{-1}\ln L/\ell$
for $d=2$, i.e.\ that the correction to the classical behaviour,
$\beta(g)=0$, was indeed negative. Assuming also the
renormalisibility, supported by the cancellation of all the
corrections to $g$ of order $[g_0^{-1}\ln L/\ell]^2$ also found in
\Cite{GLKh}, would support the large-$g$ limit of the
one-parameter picture on Fig.~\ref{beta} which, however, has been
confirmed directly only within the \N. The result
$\beta(g)=-1/g+o(1/g^2)$ found in [2--5] 
means that all the main logarithmic contributions, $[g_0^{-1}\ln
L/\ell]^n$, should be cancelled in the perturbation theory, while
diagrammatically such a cancellation has only been demonstrated
\cite{GLKh} for $n=2$.

\begin{figure}[b]\vspace*{6mm}
   \centerline{\includegraphics[width=12cm,angle=0]{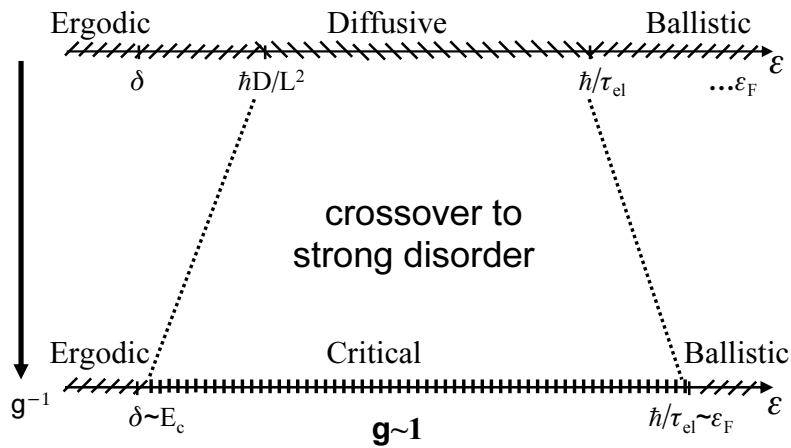}}
   \caption{Different regimes for disordered systems: the ergodic regime corresponds
   to the times longer than $L^2/D$, the time of
   diffusion through the sample, while the ballistic to the times shorter than
   $\tel$; at the critical disorder, $g\sim1$, the former regime shrinks to the
   energies of order or smaller than the level spacing $\delta$
   and the later to the energies of order of $\ef$, while the diffusive regimes
   evolves to a critical one embracing all the energies between $\delta$ and $\ef$
   \cite{ArKL}.  }
 \label{regimes} \end{figure}

Having been originally designed for describing the electron motion
in the diffusive regime, the \N\ has later been  generalised both
for a treatment of the (non-pertur\-bative) ergodic regime
\cite{Ef:82,VWZ}, where it was used to justify the applicability
of the random matrix theory for the description of electrons in
small metallic grains, as had been  suggested earlier \cite{GE},
and for a new approach to the ballistic chaotic regime
\cite{KhM,MirlinReview,A3S,ASO}. Thus, the \N\ covers all the
variety of conditions in electronic disordered systems (see
Fig.~\ref{regimes}). Moreover, this an indispensable analytical
tool to describe changes in properties of the disordered system
approaching, with decreasing $g$, the region of the
metal-insulator transition in $d>2$ or of the crossover to the
strong localisation in $d=2$.

Further extensions of the \N\ have been related to the emergence
of mesoscopics
\cite{Al:85,L+S:85,MExp:1,MExp:3}. When
it became clear that fluctuations of practically all observable
quantities in sub-micron and nano-samples are essential so that
the ensemble-averaged quantities alone are not sufficient for a
full characterisation of disordered systems, the \N\ turned out to
be a natural tool for finding entire distributions of different
observable quantities 
\cite{AKL:86,KhM,EF1,MirlinReview}, and describing their change
with increasing disorder \cite{AKL:86}.

The \N\ has also been extended to include interactions
\cite{Fin,CdCKLM:84,BeKi,Fin:87,CdCKL:98,AnKam:99,Lud:98,FLS:00,YL:01}
which allowed  to reproduce earlier perturbative results
 \cite{AA1,LR}  with a proper account
for the Landau Fermi-liquid parameters, and to derive the
renormalization group equations describing (at least at a
qualitative level) a metal--insulator transition in disordered
interacting systems. The \N\ has  been applied also to
superconducting systems \cite{Fin:87,FLS:00,YL:01}, in particular
to include effects of Coulomb interaction, e.g., lowering the
transition temperature $T_c$ by disorder -- in a (well understood)
deviation from the Anderson theorem \cite{And:59}; and also to
describe inhomogeneous or granular superconducting systems.
 The interest in this approach has been greatly
enhanced by the recent discovery \cite{Krav:95} of an apparent
metal-insulator transition in 2D disordered systems in zero
magnetic field.

The starting point for all further considerations will be the TOE
model given below. First I introduce a functional integral
representation for a simplest variant of the model that does not
include interactions; then I will show how to derive the \N\ from
such a representation by excluding ``fast'' degrees of freedom;
after that I will extend the \N\ by including the Coulomb and
pairing interactions.

\section{From the TOE model to a functional integral}
Here TOE stands for ``Theory of Everything''. In contrast to the
high energy physics, there is no problem to write down the TOE
Hamiltonian in context of electrons in solids:
\begin{equation}
\label{TOE}
  \hat H=\sum_i\frac{\hat p_i^2}{2m}+\sum_i V_i +\tfrac12
  \sum_{ij}V_{ij}\,.
\end{equation}
Here the second term represents schematically both lattice and
disorder potential, while the last represents all possible
two-particle interactions (for example, the Coulomb repulsion or
the Cooper attraction). In order to be really the TOE model, the
above Hamiltonian should also include spin terms, as later in this
paper in relation to superconductivity. Naturally, either with or
without spin terms, there is no hope for a rigourous approach to,
let alone the exact solution of, this model for an arbitrary
disorder/interaction strength. Moreover, using this model for many
problems in strongly correlated systems is no more fruitful than
starting entirely {\it ab initio}. However, when the aim is to
describe low-temperature properties of interacting electrons in
disorder environment, which are governed by low-energy modes only,
the TOE model is perfectly adequate and one does not need to go to
a next hierarchial level. This does not make the model easily
treatable, let alone straightforward. It became clear about 20
years ago that the interplay of disorder and interactions makes
disordered conductors drastically different from pure metals,
where the interactions reveal itself only in renormalisation of
Fermi liquid parameters. In the perturbative regime the
electron-electron interaction leads to nontrivial corrections to
both transport coefficients and thermodynamic properties. These
corrections govern numerous physical effects, like zero-bias
anomaly or negative anomalous magnetoresistance, which have been
summarised, e.g.,  in reviews \cite{AA1,LR}. Note than in the
present context a small perturbation parameter is $g^{-1}\ll1$,
independently of a value of the interaction parameter $e^2/\hbar
\vf$: in contrast to pure metals the condition of high density
($\lf\ll\aB\equiv\hbar^2/me^2$) is not necessary for a diffusive
weakly disordered system, when $\lf, \aB\ll\ell\ll L$. A set of
interesting questions is related to behaviour of the system when
the disorder is increasing.

The most consistent treatment of the TOE model which includes
describing its properties with increasing disorder can be achieved
within the \N\ that can be derived from Hamiltonian \Ref{TOE} for
any reasonable interaction $V_{ij}$, including both Coulomb and
pairing interactions. However, disorder makes the model nontrivial
even in the absence of any interaction. So the first task here is
a derivation of the \N\ in the most straightforward case of
$V_{ij}=0$. A non-trivial step in this derivation would be the
ensemble averaging  over all realisations of disorder. But before
coming to that, it is necessary to represent Green's functions as
functional integral.

\subsection{Introducing Green's functions}
Any physical quantity (any response function) can be represented,
at least in the absence of interaction, as a product of
one-particle Green's functions.  To this end, let's first write
down the standard formal representation for the one-particle
Green's function $\hat G(\varepsilon)$, defined by
$(\varepsilon-H)\hat G(\varepsilon)=\openone .$
 For $V_{ij}=0$, one formally
diagonalize the Hamiltonian as $   \hat H\ket \alpha
=\varepsilon_{ \alpha}\ket \alpha .$ Using the completeness
condition, $\sum_\alpha\ket \alpha\bra\alpha=\openone$, one has
(with $\hbar=1$ from now on)
\begin{equation}\label{GF}
   \hat G^{{\pm}}\equiv\Lr(){\varepsilon-\hat H{\pm i\delta}}{-1} =
 \Lr(){\varepsilon-\hat H{\pm i\delta}}{-1}\sum_\alpha\ket\alpha\bra\alpha
 =\sum_\alpha\frac{\ket\alpha\bra\alpha}{\:\;\varepsilon-\varepsilon_{ {\alpha}}
  {\pm i\delta}}
\end{equation}
where infinitesimal $\delta$ is introduced to avoid divergences at
$\varepsilon=\varepsilon_\alpha$. On recognises that $\hat G^\pm$
are the standard retarded and advanced Green's functions
\cite{AGD} which can be re-written in the real space
representation as
\begin{equation}\label{GFr}
     G^\pm_\varepsilon(\b r,\b r')\equiv \bra {\b r}\hat G^\pm\ket {\b r'}=
\sum_\alpha\frac{\braket{\b r} \alpha\braket\alpha{\b r'}
}{\varepsilon-\varepsilon_{ {\alpha}}
  {\pm i\delta}}\equiv\sum_\alpha\frac{\varphi^*_\alpha\lr(){\b r}
  \varphi_\alpha(\b r')
}{\varepsilon^\pm-\varepsilon_{ {\alpha}}
  }
\end{equation}
In the presence of disorder, one does not know either
eigenfunctions, $\varphi_\alpha(\b r)$,  or eigenvalues,
$\varepsilon_{ {\alpha}}$, of $\hat H$. The task is to represent
the average products of $G$ without any reference to unknown
parameters. To this end, I shall first represent $G$ as a
``functional integral'' which is just a multiple product of
Gaussian integrals.

\subsection{From Gaussian to functional integrals}
The first step is representing the denominator of Green's function
as a Gaussian integral. Noting that (with $\varepsilon^+\equiv
\varepsilon+i\delta$)
$$
I_0=\int_{-\infty}^{\infty}\!\!\dd
x\,\ee^{i(\varepsilon^+-\varepsilon_\alpha)x^2}
 =\sqrt{\frac\pi{i(\varepsilon^+\!-\!\varepsilon_\alpha)}}\,,
$$
one represents $(\varepsilon^+\!-\!\varepsilon_\alpha)^{-1}$ as
$(i/\pi)I_0^2$. (One assumes from now on that energy is
dimensionless, which will make all the ``action'' functionals
below also dimensionless).  Similar representation would be, of
course, possible for $(\varepsilon^-\!-\!\varepsilon_\alpha)^{-1}$
but not just for $(\varepsilon \!-\!\varepsilon_\alpha)^{-1}$: the
convergence in $I_0$ is ensured by the presence of $i\delta$.
Representing $I_0^2$ as a product of integrals over $\dd x$ and
 $\dd y$ and changing variables to $c=x+iy$ and $c^*=x-iy$,
one finds
\begin{equation}\label{Gaus}
 \frac1{\varepsilon^+\!-\!\varepsilon_\alpha}=\int\frac{\dd c^*\dd
 c}{2\pi}\,\ee^{ic^*(\varepsilon^+-\varepsilon_\alpha)c}\equiv
 Z_0\,.
\end{equation}
This gives the following representation of $\det
(\varepsilon^+-H)^{-1}=\prod_\alpha(\varepsilon^+-\varepsilon_\alpha)^{-1}$
as a product of Gaussian integrals:
\begin{equation}\label{det}
    \det(\varepsilon-H)^{-1}=\prod_{\alpha}\int\frac{\dd c^*_{\alpha}\dd
 c^{\phantom*}_{\alpha}}{2\pi}\,\ee^{ic^*_{ \alpha}
 (\varepsilon^+-\varepsilon_{ \alpha})c^{\phantom*}_{\alpha}}\equiv \Int\D
 c^*\D
 c\,\ee
 ^{i\sum_\alpha{c^*_{ \alpha}
(\varepsilon^+-\varepsilon_{ \alpha})c^{\phantom*}_{\alpha}}},
\end{equation}
where $\D c^*\D c$ is a symbolic notation for the measure of
integration, which is just a product over all $ \dd
c^*_{\alpha}\dd c^{\phantom*}_{\alpha}$. Of course, the above
determinant is defined as a product of an infinite number of
(well-defined) integrals, and as such, it may appear to be
ill-defined. However, if one does not pursue mathematical rigour,
this is quite a convenient representation which will be used as a
building block for that for Green's functions, the latter will be
free from trivial divergences that may appear in calculating
$\det(\varepsilon-\hat H)$.

Let us represent (exactly) the sum in the exponent in \Eq{det} by
an integral. Using the orthonormality condition for the
eigenfunctions of $\hat H$, one obtains
\begin{multline}
{\sum_\alpha{c^*_{\alpha} (\varepsilon^+-\varepsilon_{
\alpha})c^{\phantom*}_{\alpha}}}=\sum_{\alpha\beta} {c^*_{ \alpha}
(\varepsilon^+-\varepsilon_{ \alpha})c^{\phantom*}_{
\beta}}\ir{\,} \varphi_{
\alpha}^*(\bm r) \varphi^{\phantom*}_{  \beta}(\bm r)  \\= %
\ir{} \ub{ \equiv \psi^*(\bm r)} {
\sum_\alpha{c^*_{\alpha}}\varphi_{\alpha}^*
}\lr(){\varepsilon^+-\hat H}\ub{ \equiv \psi (\bm r)} {
\sum_\beta{c^{\phantom*}_{ \beta}}\varphi^{\phantom*}_{ \beta}
}\equiv \ir{}\, \psi^*(\bm r)\lr(){\varepsilon^+-\hat H}{\psi (\bm
r)}\label{psi}
\end{multline}
As $c$ and $c^*$ were variables of integration which took all
possible values in the complex plane, the complex conjugate fields
$\psi$ and $\psi^*$ take all the possible values for any argument
$\bm r$. Using the identity \Ref{psi}, the determinant of \Eq{det}
can be represented as the following functional integral over the
fields $\psi^*$ and $\psi$
\begin{equation}\label{Z}
    Z_1\equiv \det\lr(){\varepsilon^+-\hat
H}^{-1}=\Int\D\psi^*\D\psi\,\ee^{iS_+}\,,\qquad S_+\equiv \ir{}\,
\psi^*(\bm r)\lr(){\varepsilon^+-\hat H}{\psi (\bm r)}
\end{equation}
(An index in $Z_1$ is introduced here to make some notations in
what follows more consistent). The measure of integration is
written here symbolically as $\D\psi^*\D\psi $ in understanding
that, using \Eq{psi}, one can always reduce the $\psi$-integral to
a (relatively) well defined integral over $c$ and $c^*$. To
illustrate how it works I shall substitute this integration
measure and the definition of fields $\psi$ and $\psi^*$,
\Eq{psi}, into the functional average
$-i\av{\psi^*\psi}_+\equiv\frac1{Z_1}\int\psi^*\psi\,\ee^{iS_+}
\D\psi^*\D\psi $, thus  demonstrating that this functional average
provides a representation for the Green  function of \Eq{GFr}:
 \begin{align*}
-i\av{\psi^*\psi}_+&= -\frac i{Z_1}\sum_{\alpha \alpha'}
\varphi^*_{\alpha'}(\b r') \varphi^{\,}_{\alpha}(\b r)
\ub{=0\text{\ unless \ } \alpha=\alpha'} {\int \prod_\beta
\frac{\dd c^*_\beta \dd c^{\,}_\beta}{2\pi
i}\,c^*_{\alpha'}c^{\,}_{\alpha}\ee^ {ic^*_\beta
\lr(){\varepsilon^+-\varepsilon_{\beta} } c^{\,}_\beta}}
\\&=
-\frac i{Z_1}\sum_{\alpha}\varphi^*_{\alpha}(\b r')
\varphi^{\,}_{\alpha}(\b r) \!\!\int \frac{\dd c^*_\alpha \dd
c^{\,}_\alpha}{2\pi i}\,c^*_{\alpha}c^{\,}_{\alpha} \ee^
{ic^*_\alpha \lr(){\varepsilon^+-\varepsilon_{\alpha} }
c^{\,}_\alpha} \!\!\prod_{\beta\ne\alpha} \frac{\dd c^*_\beta \dd
c^{\,}_\beta}{2\pi i}\,\ee^ {ic^*_\beta
\lr(){\varepsilon^+-\varepsilon_{\beta}
} c^{\,}_\beta}\\
&= \frac1{Z_1}\sum_\alpha\frac{\varphi^*_{\alpha}(\b r')
\varphi^{\,}_{\alpha}(\b r)}
{\lr(){\varepsilon^+-\varepsilon_{\alpha} }^2}
\prod_{\beta\ne\alpha} \frac1{\varepsilon^+-\varepsilon_{\beta} }=
\frac1{Z_1}\sum_\alpha\frac{\varphi^*_{\alpha}(\b r')
\varphi^{\,}_{\alpha}(\b r)} { {\varepsilon^+-\varepsilon_{\alpha}
}} \ub{\equiv {Z_1}}{\prod_{\text{all }\,\beta}
\frac1{\varepsilon^+-\varepsilon_{\beta} }}\\
 &=\sum_\alpha\frac{\varphi^*_{\alpha}(\b r')
\varphi^{\,}_{\alpha}(\b r)} { {\varepsilon^+-\varepsilon_{\alpha}
}} \equiv G^+_\varepsilon(\b r, \b r')\,.
\end{align*}
However, the functional integral would be useless, had one every
time needed to return to the $c$ representation in order to make a
calculation. It is sufficient to realise once and forever that the
functional integral is a short-hand notation for a product of
Gaussian integrals, so that all the standard operations for
Gaussian integrals can be directly translated into the language of
functional integrals. To illustrate such a translation, let us
note that in the above calculation the following elementary
integral has been used:
$$
 I_0\equiv\int \frac{\dd c^*_\alpha \dd
c^{\,}_\alpha}{2\pi i}\,c^*_{\alpha}c^{\,}_{\alpha} \ee^
{ic^*_\alpha \lr(){\varepsilon^+-\varepsilon_{\alpha}}}=\frac
i{(\varepsilon^+-\varepsilon_{\alpha})^2}\equiv\frac
i{\varepsilon^+-\varepsilon_{\alpha}}Z_0\,,
$$
where $Z_0$ is the integral \Ref{Gaus}. There are many ways to get
such a straightforward result. Let us choose that which is
directly generalisable to the functional integral, representing
$I_0$ above using parametrical derivatives with respect to the
``source fields'' $h$ and $h^*$
\begin{align}
\notag I_0&=\frac1{Z_0}\int \frac{\dd c^*\dd c^{\,}}{2\pi
i}\,c^*c\,\ee^ {ic^*Ac}= -\frac1{Z_0}\diff {h^*}h\int \frac{\dd
c^*\dd c^{\,}}{2\pi i} \,\ee^ {i\lr(){c^*Ac+h^*c+c^*h}}\\&=
-\frac1{Z_0}\diff {h^*}h\int \frac{\dd c^*\dd c^{\,}}{2\pi i}
\,\ee^ {i\lr(){c^*Ac-h^*A^{-1}h}}=-\diff {h^*}h
\ee^{-ih^*A^{-1}h}=\frac iA\,,\label {I}
\end{align}
where a shift of variables, $c\to c-A^{-1}h$ and $c^*\to
c^*-h^*A^{-1}$, has been used. It was implied that both $h$ and
$h^*$ are put to zero after calculating the derivatives above. Now
I apply practically the same procedure to the functional integral
defined above in order to represent Green's functions.

\subsection{Green's function as a functional integral} To
show directly that $-i\av{\psi^*\psi}_+$ represents Green's
function, I express it as a functional derivative:
\begin{align}\notag
-i\av{\psi^*(\bm r_1)\psi(\bm r_2)}_+&=\frac i{Z_1} \dff
{}{h^*(\bm r_1) }{h(\bm r_2)}\psiint \ee^{iS_+(h,h^*)}\\[-8pt]\label{FI}\\[-8pt]
S_+(h,h^*)&\equiv \ir{\lr[]{ \psi^*(\bm r)\lr(){\varepsilon^+-\hat
H}{\psi (\bm r)}+ h^*(\bm r)\psi (\bm r) +\psi^* (\bm r) h(\bm
r)}}\,.\notag
\end{align}
Noting that, by definition, $\lr(){\varepsilon^+-\hat
H}^{-1}=G^+_\varepsilon$ and making a shift of variables similar
to that above, $\psi \to \psi -G\cdot h$ and $\psi ^*\to \psi
^*-h^*\cdot G$ (with $(G\cdot h)(\bm r)$ being a convolution in
the $\bm r$ space) one reduces $ S_+(h,h^*)$ in the functional
integral to the following expression:
$$
 S_+(h,h^*)\to S_+-\ir{\lr[]{h^*(\bm r) G^+_\varepsilon(\bm r, \bm
 r')h(\bm r')}}\dd^d r'\,,
$$
so that performing the functional differentiation one finds,
indeed, that
\begin{align}\label{GFF}
G^+_\varepsilon(\bm r_1, \bm r_2)&=-\frac i{Z_1}\psiint\psi^*(\bm
r_1)\psi(\bm r_2)\ee^{iS_+}=\frac1{Z_1} { \dff{{Z_1}(h^*,h)}
{h^*(\bm r_1)} {h(\bm r_2)}}{\lower
5pt\hbox{$\Bigg|$}}{}_{{\begin{matrix}
_{h^{\phantom *}=0}\\[-4pt]
_{h^*=0}\end{matrix}}}\\
{Z_1}(h^*,h)&\equiv\psiint\ee^{iS(h^*,h)}\,.\notag
\end{align}
That completes the representation of Green's function as a
functional integral. If one thinks that the above derivation was
not ``pedestrian'' enough, it would be sufficient to compare it to
the standard way of introducing functional integrals (see, e.g.,
\Cite{NegOrl}). Note that \Ref{FI} is not an action as usually
understood: in the absence of time integral, it does not even have
the dimensionality of $\hbar$, and is simply assumed
dimensionless. Unfortunately, such a simplified way  is only
possible for a quadratic Hamiltonian which can be
straightforwardly diagonalized. Only  the presence of disorder
makes it nontrivial: $\hat H$ in \Eq{Z} contains a random
potential $V$. The ensemble averaging over the distribution of $V$
would result in an effective functional containing higher powers
of $\psi^*$ and $\psi$. If $\hat H$ is nonlinear to begin with,
like the Hamiltonian of the TOE model \Ref{TOE}, one should choose
a hard way \cite{NegOrl} to get a functional representation for
Green's functions. However, there is an alternative. After having
learned a bit how to deal with the functional integrals that
emerged for the  noninteracting problem, one would be able to use
some leap of faith to extend considerations to include
interactions. Such an approach would not be self-contained but
would remain, in my opinion, reasonably consistent and believable.

A certain extension related to future considerations is useful
right now. Particle statistics are totally irrelevant for
one-particle problems, but in extending considerations to the
interacting particles one shall be mainly interested  in fermions.
The functional integration described above could be extended to
the case  of interacting bosons rather than fermions as the
commuting classical fields in the functional integrals correspond
to operators with certain commutation relations in the Hamiltonian
representation. As fermions are described by operators with
certain anticommutation relations, it is not that surprising that
the functional integral representation for fermions must be
formulated in terms of anticommuting classical fields. There
exists a very solid mathematical foundation of the theory of
anticommuting classical variables \cite{Berezin} reduced to a
physical level of rigour in different derivations \cite
{EfLKh,Ef:82,AKL:86} of the \N. However, all one needs for the
purpose of present considerations is a relatively moderate leap of
faith that allows one to introduce a fermion representation  at a
one-particle level.

\subsection{Fermions do it like bosons} but upside down: instead
 of having $\det^{-1}$ in the partition function for bosons \Ref
Z, one has $\det$ in the appropriate fermionic representation:
\begin{align}\label{ZF}
{Z_1}^+\equiv \det\lr(){\varepsilon^+-\hat H} &=\Int\D{\overline
\psi}\D\psi\,\ee^{iS_+}\,,\qquad S_+\equiv \ir{\lr[]{ {\overline
\psi}(\bm r)\lr(){\varepsilon^+-\hat H}{\psi (\bm r)}}}\,,\\
G^+_\varepsilon(\bm r_1, \bm r_2)&=-\frac
i{Z_1^+}\Psiint\overline\psi(\bm r_1)\psi(\bm
r_2)\ee^{iS_+}\label{GF+}
\end{align}
where $\overline \psi$ and $\psi$ are classical anticommuting
mutually conjugate fields. The conjugation requires some care. For
the bosonic classical fields it was implied that $\psi,\psi^*
=a\pm ib$, with $a$ and $b$ being real fields. For the fermionic
fields it is more convenient to introduce the conjugation
operation in matrix notations:
\begin{equation}\label{conjugate}
\psi= \frac{1}{\sqrt 2}
\begin{pmatrix}\varkappa\cr{\varkappa^*}\end{pmatrix},\qquad {\overline \psi}=
\frac{1}{\sqrt 2}\,\bigl(
-{\varkappa^*}\quad\varkappa\,\bigr)\,,\qquad\Lr(){\varkappa^*}*=-\varkappa\,,\qquad
{\overline \psi}\equiv\Lr(){{\mathcal C}\psi}{\text T}
\end{equation}
The last equation above defines the so called
``charge-conjugation'' matrix ${\mathcal C}$ (which has nothing to
do with the electric charge).  The symmetry imposed by such a
conjugation will not be used but in section \ref{NF}. In what
follows, it is sufficient to know that the classical fermionic
fields anticommute (which implies not only
$\varkappa{\varkappa^*}=-{\varkappa^*}\varkappa$ but also
$\varkappa^2={\varkappa^*}^2=0$) and the partition function
\Ref{ZF} is given by $\det(\varepsilon-\hat H)$ rather than by
$\det^{-1}(\varepsilon-\hat H)$.

Now everything is ready for the ensemble averaging over disorder
and the only question to ask is what to average. The averaged
one-particle Green's function $\av G$ is not an interesting
object: in the absence of interactions disorder does not effect
physical quantities which are directly expressible via $\av G$,
such as density of states. Most of response functions of interest
such as density-density correlations, conductance, mesoscopic
fluctuations of any observable, etc., can be normally expressed
via the averaged products of retarded and advanced Green's
functions. So the simplest product to average over disorder is
$G^+_\varepsilon G^-_{\varepsilon-\omega}$, with $G^+$ represented
by \Eq{GF+} and $G^-$ represented as
\begin{align}
G^-_{\varepsilon-\omega}(\bm r_1, \bm r_2)&=-\frac
i{Z_1^-}\Psiint\overline\psi(\bm r_1)\psi(\bm
r_2)\,\ee^{iS_-}\label{GF-}\,,
\end{align}
where $S_-$ differs from $S_+$ in \Eq{ZF} by substituting
$\varepsilon-\omega^+\equiv\varepsilon-(\omega+i\delta)$ for
$\varepsilon^+\equiv \varepsilon+i\delta$. In general, one
averages a closed loop, i.e.\ a product like $G(\bm r_1, \bm
r_2)G(\bm r_2, \bm r_3)\ldots G(\bm r_n, \bm r_1)$. This allows
one to introduce quite an economical representation, using one
source field for a product $\psi(\bm r)\opsi(\bm r)$ of two fields
taken from the adjacent Green's function:
\begin{align}\label {GG}
K_\omega(\bm r, \bm r')\equiv G^+_{\varepsilon+\omega^+/2}(\bm r,
\bm r') G^-_{\varepsilon-\omega^+/2}( \bm r', \bm r)&= \dff{\ln
{Z_1}(\bm \eta)}{\eta_{12}(\bm r)}
{\eta_{21}(\bm r')} {\lower 3pt\hbox{$\bigg|$}}_{\bm \eta=0}\\
{Z_1}(\bm
\eta,\omega)&=\bpsiint\ee^{iS_1(\bm\eta,\omega)}\notag\\
S_1(\bm\eta,\omega)\equiv\int\biggl\{\overline{\bm \psi}(\bm
r'')\biggl(-\hat \xi  -V+\tfrac12\omega^+&\Lambda\biggr) {\bm
\psi}(\bm r'')+\lr(){\overline{\bm \psi}\bm\eta\bm \psi}_{\bm r''}
\biggr\}\dd^dr''\,.\label{S}
\end{align}
Here $\hat \xi \equiv \hat {\bm p}^2/2m-\ef$; the frequency
arguments have been shifted for convenience;  $\overline {\bm
\psi}\equiv \lr(){\opsip, \opsim}$, with $\opsi^\pm$ having been
used in representations for $G^{\pm}$, respectively; ${\bm \psi}$
is a corresponding vector column; $\bm \eta$ is an arbitrary
$4\times4$ matrix field (whose diagonal elements, $\eta^{++}$ and
$\eta^{--}$, are redundant and can be put to zero from the very
beginning, if convenient, while $\eta^{+-}$ and $\eta^{-+}$ are
$2\times2$ matrices by themselves, as each $\psi^\pm$ is a
two-component field, \Eq{conjugate}); the constant matrix
$\Lambda\equiv \diag (1,1,-1, -1)$ is necessary as $\omega^+$
enters with the opposite sign in the actions $S^\pm$; ${Z_1}$ is
the functional integral in the r.h.s.\ of \Eq{GG} calculated at
$\bm \eta=0$. As the $\eta$-part of the action is not affected by
the averaging, the resulting functional would be applicable to any
product of $G^+$ and $G^-$ subject to a proper extension of the
matrix space (each $G$ would require its own pair of fields in the
functional representation) and an appropriate choice of the source
fields $\eta$.

\section{Averaging over disorder}
For simplicity, one assumes the random potential $V(\bm r)$ in
\EQS {TOE}S to be a Gaussian white noise with $\av{V(\bm r)}=0$
and
\begin{align}
\label{VV}%
\av{V(\bm r) V(\bm r')}=\frac1{2\pi \nu_0 \tel}\delta(\b r-\b
r')\,,
\end{align}
which means that $V(\bm r)$ is taken from the following
distribution:
\begin{align}
\label{PV} P\lr\{\}{V(\bm r)}&={\mathcal N}
\exp\lr[]{-\pi\nu_0\tel\ir{ V^2(\bm r)}}\,,
\end{align}
where $\mathcal N$ is a normalisation factor. Although the choice
of a white-noise Gaussian potential makes calculations
considerably easier, a generalisation to a slowly varying
potential would not basically change results
\cite{ASO,GornMirlWolf:01} provided that the mean free path $\ell$
is substituted by the transport length $\ell_{\text{tr}}$. With
this choice, the disorder averaging $\av{\ldots}$ reduces to
taking a functional average with the distribution \Ref{PV}. Had
one to average ${Z_1}$, this would be a straightforward Gaussian
integral again. It is not straightforward to average $\ln {Z_1}$,
though. If one first takes derivatives with respect to the source
fields in \Eq{GG}, it is the presence of the denominator,
$1/{Z_1}(\eta\!=\!0)$, that makes the averaging tricky.

This denominator is the partition function  given by \Eq Z in the
bosonic and by \Eq{ZF} in the fermionic representation. Comparing
these expressions, one can see that using a mixed,
\textit{supersymmetric } fermion-boson representation would cancel
out this denominator. This idea \cite{Ef:82} turned out to be
extremely fruitful and led to enormous progress in theory of
\textit{non-interacting} particles in disordered and chaotic
systems. Numerous achievements of the supersymmetry method have
been summarised in a book \cite{Efetbook} and a recent review
\cite{MirlinReview}. However, there is no simple way of applying
this method to a many-particle system of {\it interacting}
electrons, so that some other method is required.

After a few earlier attempts
\cite{ArIos,BabKoz,HorSch},
it has recently been  demonstrated \cite{AnKam:99,Lud:98} that the
Keldysh technique \cite{Keld,RS:86} provides a viable route to a a
field-theoretical description. In this technique  $Z=1$ by
construction, which makes the ensemble-averaging straightforward.
However, this technique remains quite cumbersome and, although it
might be indispensable in describing non-equilibrium problems, a
considerably simpler alternative exists for interacting electrons
in disordered systems at equilibrium (i.e.\ in the linear response
regime for all transport problems).

This alternative is a so called replica trick, which is possibly
the oldest method \cite{EdAn,Em} used for the disorder averaging.
It is based on the following identity:
 \begin{align}\label{n}
\ln {Z_1}=\lim_{n\to0}\frac{{Z_1}^n-1}n\,.
 \end{align}
To apply it one \textit{replicates} the fields in \Eq{GG},
$\bm{\opsi}\mapsto {\overline \Psi}
\equiv(\opsip_1,\ldots,\opsip_n, \opsim_{1},\ldots,\opsim_{n}$),
thus  replacing it by the following:
\begin{align}\label {K}
K_\omega(\bm r, \bm r') =
\lim_{n\to0}\frac1{2n^2}\tr\!\left\{\dff{ {Z}(\bm
\eta)}{\bm{\overline\eta} (\bm r)} {\bm\eta (\bm
r')}\right\}_{\!\bm \eta=0}  \,,\quad Z\equiv Z_n=Z_1^n
=\bPsiint\ee^{iS_n(\bm\eta)}\,.
\end{align}
The functional integration in \Eq K implies a product of $n$
identical integrals; $S_n$ is given by the same expression \Ref S
as $S_1$ but in terms of the replicated fields and with
$\Lambda\equiv \diag (\openone_{2n}, -\openone_{2n})$ being a
block-diagonal $4n\times4n$ matrix; $\bm \eta$ is a
block-off-diagonal $4n\times4n$ matrix with
$\bm\eta^{++}=\bm\eta^{--}=0_n$ and $\bm\eta^{+-}=\bm\eta^{-+}$
being arbitrary $2n\times 2n$ matrices:
\begin{equation}\label{SR}
S(\bm\eta)\equiv\ir{\biggl\{\Psib(\bm r)\biggl[-\hat \xi  -V(\bm
r)+\tfrac12\omega^+\Lambda+\bm \eta\biggr] \bPsi(\bm r)
\biggr\}}\,.
\end{equation}
An  extra factor of $1/2n$ in \Eq K compared with \Eq n is due to
the fact that the trace there gives the sum of $2n$ identical
terms. From now on I will be dealing only with the replicated $Z$
and $S$ and thus omit the index $n$ there.

As the representation \Ref K does not contain a logarithm, one
only needs to average $Z$ over the disorder distribution of
\Eq{PV}, which, for any {\it integer} $n$, is a straightforward
Gaussian integration that involves only the $V$-dependent part of
$Z$:
\begin{align}
    \Int\frac{\D V}{\mathcal N} \ee^{\mbox{$\int$}{-\lr[]{\pi\nu_0\tel V^2+i\Psib
 V {\Psi}}}\dd^dr}=\exp\lr[]{-\frac{1}{4\pi\nu_0\tel}\ir{ {\Lr(){\Psib\Psi}
 2}}}\,.\label{G}
\end{align}
Thus one arrives at the following effective action:
\begin{align}
S (\bm\eta)\equiv\int\biggl\{\overline{\Psi}(\bm r)\biggl(-\hat
\xi +\tfrac12\omega^+&\Lambda+\bm\eta\biggr) {\Psi}(\bm r)+
\frac{i} {4\pi\nu_0\tel}  \Lr(){ \Psib\Psi}2
\biggr\}\dd^dr\,.\label{s}
\end{align}
After calculating $K_\omega(\bm r, \bm r')$ for any positive
integer $n$ one needs to take the replica limit $n\to0$ which is
by no means a well defined mathematical operation.

\begin{figure}[t]
  \begin{center}{
  \scalebox{.65}{\includegraphics*[clip,angle=0]{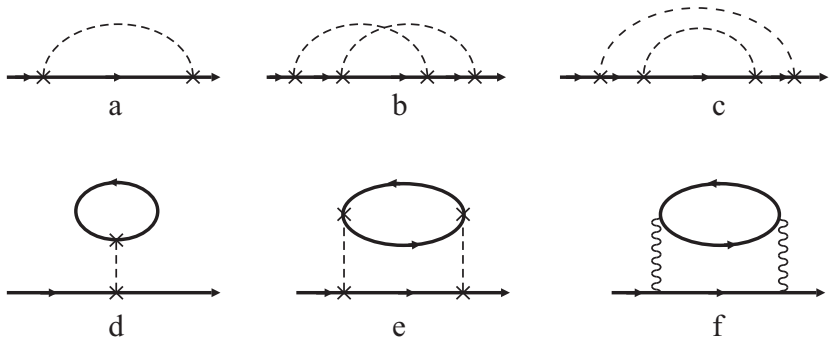}}}
   \end{center}
   \vspace{-5mm}

\caption{Diagrammatic corrections for one-particle Green's
function:
   dashed lines correspond to pair correlations of the random potential,
   \Eq{VV}, or equivalently to the ``interaction'' (the bilinear term)
   in \Eq s; wavy lines correspond to a real interaction introduced in
   section \ref{interactions}.}
\label{diagrams} \end{figure}

However, as long as one is interested in perturbative corrections
only (including the RG corrections), one can talk about the
replica method, not a trick, as illustrated in
Fig.~\ref{diagrams}. All the perturbative corrections in the
standard diagrammatic techniques are just crosses (corresponding
to scattering from impurities) connected by solid lines
corresponding to Green's function, like diagrams (a) -- (c), where
the dashed lines connecting  the crosses represent the pair
correlations of the random potential \Ref{VV} resulting from the
ensemble averaging. The field theory with the functional \Ref s
with any integer $n$ exactly reproduces all such diagrams but also
yields those like (d) and (e) with vacuum loops. However, each
loop contains a sum of $n$ identical replicated contributions
which lead to an extra factor of $n$ thus making such diagrams to
vanish in the replica limit $n\to0$. Note that in the
supersymmetric technique such diagrams vanish as each loop is the
sum of fermionic and bosonic contributions having the opposite
signs. However, in the latter approach it would be difficult to
include diagrams like (f) that contains a loop corresponding to
the electron-hole excitation in the presence of the interaction,
whereas the replica approach would encompass such contributions
rather naturally (section \ref{interactions}).

The functional of \Eq s is fully equivalent to the original model
of \Eq {TOE} in the absence of interactions. However, there is
little to gain from it without further approximations. The most
important one is based on a reduced description that excludes all
details of electronic motion at scales smaller than the mean free
path $\ell$.

\section{The \N\ functional}
\label{NF}

\subsection{Hubbard--Stratonovich
decoupling} This is the name of the operation performed in order
to exclude all ``fast'' modes of electron motion. Its first step
is nothing more than the Gaussian integration like that in \EQS I
G, but the integration which is taken ``backwards''. In \Eq G, one
integrates over the random potential $V$ thus arriving at the
action \Ref s that does not depend on $V$ (and thus contains no
randomness) but is no longer quadratic (i.e.\ integrable) in the
fermionic fields. In order to decouple the quartic term in \Eq s,
one introduces a new field, $Q$, which allows one to ``unwind''
the integration in \Eq G and represent the quartic term as a
functional integral over $Q$, thus making the action quadratic in
the fermionic fields again and finally integrating over
$\bPsi,\,\Psib$ to arrive at the action in terms of the new field
$Q$. Of course, one could perform the $\Psi$-integration from the
very beginning obtaining the action in terms of the disorder
potential $V$; however, such an action would not be  very useful.
The introduction of the field $Q$ is useful if it exploits fully
the configurational space defined by the conjugate fields $\Psib$
and $\bPsi$ and excludes fast degree of freedom; to this end, $Q$
is chosen as a matrix field having the same rank and symmetry as
$\bPsi\otimes\Psib$ (the latter is just a notation for an outer
product of two vectors that results in a $4n\times4n$ matrix).
Noting that $\tr\Lr(){\bPsi\otimes\Psib}2=-\Lr(){\Psib\bPsi}2$ and
choosing $Q$ to be dimensionless, one can write the following
identity that ``unwinds'' the integration \Ref G:
\begin{align}
\exp\lr[]{-\frac{1}{4\pi\nu_0\tel} \ir {{\Lr(){\Psib\Psi}
 2}}}
 = \Int\frac{\D
Q}{\mathcal N_Q} \exp\left\{ \ir {\lr[]{ -
\frac{\pi\nu_0}{8\tel}\tr Q^2+\frac{i}{2\tel}{ \Psib}Q\bPsi}}
\right\}
 \,.\label{g}
\end{align}
The normalisation integral $\mathcal N_Q$ is just the same
$Q$-integral at $\Psib\!=\!\bPsi\!=\!0$ (and it is anyway
irrelevant in the replica limit) and the Gaussian integration both
here and in \Eq G is performed with the help of a shift of
variables fully analogous to those in \EQS I{FI}. After this, the
$\bPsi$-dependent part of the action is, indeed, quadratic, with
the term $\frac{i}{2\tel}{ \Psib Q\bPsi} $ being added into the
brackets in \Eq s. The Gaussian integration over the fermionic
field in \Eq g is performed exactly like in \Eq {ZF} (the rank of
matrices does not matter) and results, after using the remarkable
matrix identity $\ln\det A=\tr\ln A$, in the effective action
$\cF$ which depends only on $Q$:
\begin{eqnarray}
\label{Trln}
{\cF}[Q ]=\frac{1}{L^d}\int \text{d}^dr \left\{\frac{\pi
}{8\tau_{\text{el}}\delta_0}\tr Q ^2-\frac12\tr \ln\lr[] {
\frac{\omega} 2 \Lambda +\bm\eta-\hat \xi +\frac i
{2\tau_{\text{el}}}Q  } \right\}\,,
\end{eqnarray}
where $\delta_0\equiv1/(\nu_0 L^d)$ is the mean level spacing. The
action has been renamed into $\cF$ in order to stress that the
functional averaging in \Eq g is performed with the Euclidian
weight $\ee^{-\cF}$ rather than with the oscillating one,
$\ee^{iS}$. The partition function $Z$, \Eq K, is now given by
\begin{equation}\label{z}
 Z=\Int\D Q\exp\lr(){{-\cF}[Q ]}\,,
\end{equation}
and the measure of integration will be specified later. The matrix
$Q$ is Hermitian as it has the same symmetry as the product
$\bPsi\otimes\Psib$ which is Hermitian due to the definition of
the fields in \Eq{conjugate}. This is not the only symmetry of
$Q$, as the charge conjugation introduced in \Eq{conjugate} is
essential and explains the appearance of the factor $1/2$ before
the $\tr\ln$ in \Eq{Trln}, mysterious for generations of students.

\begin{figure}[t]
  \begin{center}{
  \scalebox{.65}{\includegraphics*[clip,angle=0]{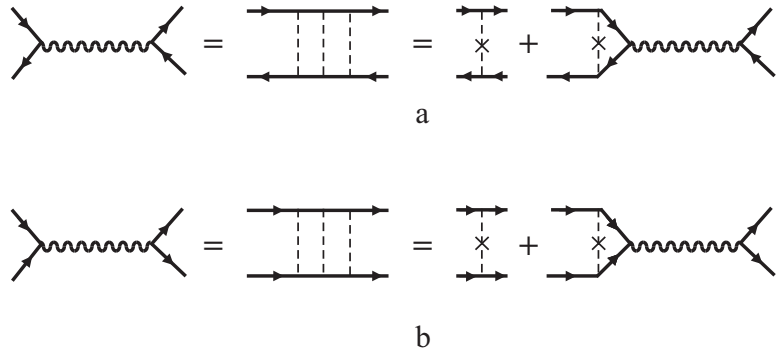}}}
   \end{center}
   \vspace{-5mm}

\caption{Diffuson (a) and cooperon (b) ladders:
   in diagrammatic techniques either pictures with
   triple dashed
   lines or those with wavy lines can be used
   to denote the ladders; the latter are more convenient in the
   present context as they are in direct correspondence with
   diagrammatic notations in the \N, section \ref{diagrammatics}; they should
   not be confused with interaction lines, Fig.~\ref{diagrams}.}
\label{diffuson} \end{figure}
If a diligent student had taken a course of action described
between \EQS g{Trln}, he would have duly arrived at \Eq{Trln} but
without this $1/2$. Such a result, moreover, would be formally
correct but with a matrix $Q$ ill suited for describing the slow
variables. The point is that the field $Q$ should be chosen  in
such a way that it would correspond only to the slow part of
$\tr\Lr(){\bPsi\otimes\Psib}2$ in the l.h.s.\ of \Eq g while the
fast Fermi oscillations contained in each field $\psi$ were
suppressed. Such a slow part could be chosen in two ways: it
corresponds either to the  $\bPsi\Psib$ pairing -- with a small
transferred momentum in the Fourier space, or to the
$\Psib\,\Psib$ (and $\bPsi\bPsi$) pairing -- with the small sum
momentum. The charge conjugation introduced in \Eq{conjugate}
allows automatically for the existence of the two pairing
channels, provided that the field $Q$ is chosen to have the same
charge-conjugation symmetry as the product $\bPsi\otimes\Psib$,
i.e.\ $Q={\mathcal C}Q^*{\mathcal C}^{-1}$. Allowing also for the
fact that $Q$ is Hermitian (which is, of course, necessary for at
least a superficial convergency of the $Q$ integral in \Eq g), one
can represent the $4n\times 4n$ matrix field $Q$ as
\begin{equation}
\label{dc}
 \displaystyle Q= \left(
\begin{array}{cc}
d & c \\
c^+ & d^T
\end{array}
\right), \qquad  \quad d^+ = d, \,\,\, c^T = -c,
\end{equation}
where $d$ and $c$ are $2n\times 2n$ matrices in the replica space,
corresponding to the decoupling of the $\bPsi$ pairs with the
small transverse or small sum momentum, respectively. Then, in
going from \Eq g to \Eq {Trln}, one integrates independently over
fields $d$ and $c$ which leads to the appearance of the factor
$1/2$ in  the latter equation. Some familiarity with the standard
diagrammatics may help one to recognise that $d$ corresponds to a
{\it diffuson} -- the sum of impurity ladder diagrams with a small
transfer momentum, while $c$ corresponds to a cooperon -- the sum
of ladder diagrams with a small sum momentum, Fig.~\ref{diffuson}.
To make it a working analogy, one needs to use the fact that in
\Eq {Trln} $Q$ is a slow varying field. To this end, one should
expand the functional \Ref{Trln} in gradients of this field.

\subsection{Gradient Expansion} The first step is to define a
class of stationary, spatially-homogeneous (i.e.\ `infinitely
slow'') fields for which the functional \Ref{Trln} satisfies a
saddle-point condition.  By varying the functional with respect to
the field $Q$, one finds
\begin{eqnarray}
\label{SP}-i\pi \nu_0 Q=\bra {\bm r}  \G\ket {\bm r} \equiv
 \bigg\langle{\bm r}\bigg|
 \Lr()
{ \frac\omega 2 \Lambda+\bm\eta -\hat \xi +\frac i
{2\tau_{\text{el}}}Q }{-1} \bigg|{\bm r}\bigg\rangle \,.
\end{eqnarray}
This saddle-point condition is obviously satisfied {\it at zero
frequency $\omega$} and in the absence of the source field $\bm
\eta$  by any spatially-homogeneous matrix $Q$ of a proper rank
and symmetry that obey the condition $Q^2=\openone$. This
condition allows one to invert explicitly the matrix in the square
brackets in \Eq{SP}: in the Fourier momentum space $\xi\equiv\bm
p^2/2m-\ef$ becomes just a scalar variable of integration so that
one finds
\begin{equation}\label{gf}
\G(\bm p,\,\omega\!=\!0)=\Lr[]{-\xi+\frac i {2\tau_{\text{el}}}Q
}{-1}= \frac1{\xi^2+\frac1{4\tel^2}}\,\lr(){\xi+\frac i
{2\tau_{\text{el}}}Q }\,.
\end{equation}
After this,  the corresponding integration over all momenta in the
r.h.s.\ of \Eq{SP} is straightforward in the pole approximation
(justified provided that the week disorder condition,
$\ef\,\tel\gg1$, is fulfilled), when one replaces this integration
by $\nu_0\int_{-\infty}^{\infty}\dd \xi$.

The condition $Q^2=\openone$ is convenient to parameterise as
\begin{equation}\label{1}
 Q=U^\dagger\Lambda U\,,
\end{equation}
where $U$ belongs to a so called {\it symplectic} group $\Sp(2n)$.
This means that $U^\dagger  U=1$, and each  $U$ is a ``real
quaternion'' matrix, i.e.\ can be represented as $2n\times 2n$
matrix with matrix elements of the structure $a+b\bm i +c\bm j
+d\bm k$, where $a,b,c,d$ are  real $n\times n$ matrices  and $\bm
i,\,\bm j,\,\bm k$ are the so called quaternion units, represented
as $i\hat{\bm\tau}$ with $\hat{\bm\tau}$ being the set of
$2\times2$ Pauli's matrices in the ``charge-conjugate'' space
defined by \Eq{conjugate}. Such a choice of $U$ preserves the
charge-conjugate symmetry of matrix $Q$, \Eq {dc}. However, each
$Q$ can be represented in the form \Ref 1 in infinitely many ways,
as the ``gauge transformation'' $U\mapsto SU$ leaves $Q$ unchanged
for any $S$ commuting with $\Lambda=\diag\lr(){\openone_n,
-\openone_n}$. Such a matrix $S$ is a block-diagonal matrix
$S=\diag\lr(){S_n, \widetilde S_n}$, with each $S_n$ belonging to
the symplectic group $\Sp(n)$, i.e.\ $S\in\Sp(n)\otimes\Sp(n)$. To
make the representation \Ref 1 single-valued, one needs to
factorise out matrices $S$ (i.e.\ to fix the gauge). As a result
of such a factorisation, one ends up with a coset space: $U\in
\Sp(2n)/\Sp(n)\otimes\Sp(n)$ for matrices $U$ that makes a
single-valued representation for $Q$ in \Eq 1 (as matrices $Q$ and
$U$ are now in one-to-one correspondence, $Q$ belongs to the
equivalent coset space, although its matrix representation would
be obviously different). A general geometric structure of such a
coset space is not difficult to understand and use (see, e.g., an
excellent book \cite{DFN} addressed specifically to physicists).
However, it won't be necessary since in what follows an explicit
single-valued parameterisations of matrices $U$ and $Q$ will be
used, which could be straightforwardly verified without any
benefit of knowledge of differential geometry.

Note that $Q=\Lambda$ is a solution to the saddle point equation
\Ref{SP} at nonzero $\omega$. What is usually not stressed is that
the saddle-point equation is satisfied not only by $\Lambda$ but
by any spatially-homogeneous diagonal matrix commuting with
$\Lambda$ with eigenvalues equal to $\pm1$. This must be taken
into account in considering solutions with replica-symmetry
breaking, the latter being the necessary conditions of obtaining
non-perturbative result within the replica approach
\cite{KamMez:99a,YL:99a}. However, the existence of this wide
class of the saddle-point solutions at $\omega\ne0$ is irrelevant
for most applications and, in particular, for the expansion of the
functional of \Eq {Trln} near the saddle point of \Eq {SP}. In
such an expansion, the transverse modes with $Q^2\ne1$ can be
neglected as, due to the first term in the functional \Ref{Trln}
they would incur a gap of order of $1/\tel\delta\gg1$ (for a
homogeneous excitation) and would be thus suppressed. Then the
first term gives just an irrelevant constant that could be
neglected (it vanishes in the replica limit as $\Tr1\propto n$ but
such a term would be irrelevant even in a finite-$n$ \SM.) Terms
with a smaller gap of order $\omega/\delta$ would appear upon the
expansion of the second, Tr ln term. Keeping them is justified as
long as $\omega\tel\ll1$ which will be assumed throughout. The
expansion of the Tr ln term will also produce gapless gradient
terms with the magnitude of order $(1/\tel\delta) (q\ell)^2$ in
the momentum space: keeping them while neglecting the $Q^2\ne1$
modes is justified as long as $q\ell\ll1$. Therefore, the two
conditions of the applicability of the standard \N\ are
\begin{align}\label{applicability}
\omega\tel&\ll1&q\ell&\ll1\,,
\end{align}
which should be taken together with the weak disorder condition,
$g\gg1$. Therefore, the standard  \N\ is applicable for the
description of the \textit{diffusive} regime, at scales $r\gg\ell$
and times $t\gg\tel$.  These conditions can be relaxed and the
NL$\sigma$M can be generalised
\cite{KhM,EF1,MirlinReview,A3S,ASO} for considering also ballistic
regime ($r\lesssim\ell$) but such a generalisation will not be
considered in the present lectures. Now the expansion of the tr ln
term in \Eq{Trln} is straightforward by making use of the
condition \Ref 1 and applying the similarity transformation
(omitting for now the source field $\bm \eta$):
\begin{align}\notag
\cF&= -\frac{1}{2}\Tr\ln\left[ -\hat\xi+  \frac{i}{2\tel}Q +
\frac{\omega^+}2\,\Lambda \right]=
-\frac{1}{2}\Tr\ln\left[-U\hat\xi U^\dagger
+\frac{i}{2\tel}\Lambda+\frac\omega2U\hat\Lambda U^\dagger\right]
 \\\label{gradexp}
  &=-\frac{1}{2}\Tr\ln\lr\{\}{\G_0^{-1} -U\lr[]{\hat\xi,
  U^\dagger}+\frac\omega2U\hat\Lambda
  U^\dagger}\\&=\const-\frac\omega4\Tr\lr(){\G_0U\Lambda
  \Ud}+\frac14\Tr\lr(){\G_0U\lr[]{\hat\xi,\Ud}\,
  \G_0U\lr[]{\hat\xi,\Ud}}\,,\notag
\end{align}
where $\G_0$ is Green's function \Ref{gf} with $Q=\Lambda$ and the
symbol $\Tr$ implies the integration over $\br$ as well as the
summation over matrix indices. The real part of the first term is
an irrelevant $n$-proportional constant, while its imaginary part
reduces to $-(i\omega \alpha/4)\Tr\lr(){\Lambda U\Lambda \Ud} =
-(i\omega \alpha/4)\Tr\lr(){\Lambda Q}$, with the coefficient of
proportionality $\alpha=\Lambda \Im\text m\,\G_0(\bm r,\bm r)=
\Lambda \Im\text m\,\ip{\G_0(\bm
p)}=\nu\int\dd\xi\Lr(){\xi^2+1}{-1}=\nu\pi\,.$ Thus, the first
term in the expansion \Ref{gradexp} gives the gradientless term in
the \N\ functional \Ref{nlsm}. In the second term in the expansion
\Ref{gradexp}, the operator $\hat \xi=(-1/2m)\nabla^2-\ef$ in the
commutator $\lr[]{\hat\xi,\Ud}$ reduces to $\vf\,\bm n\cdot\nabla$
($\bm n$ is a unit vector) so that Fourier-transforming it into
the momentum space and performing the angular integration, one
finds this term equal to
$$
\frac{\vf^2}4 \ir{\Bg[]{\bra {\bm r}  \G_0\ket {\bm r'}\lr(){\bm
n\cdot\bm \Ascr}_{\br'}\bra{\br'}\G_0\ket\br\lr(){\bm n\cdot\bm
\Ascr}_\br }}\dd^dr'= \frac{\vf^2}{4d}  \sum_{\bm p,\bm
q}{\Bg[]{\G_{0,\bm p}\bm \Ascr_{\bm q}\cdot\G_{0,\bm p+\bm
q}\bm\Ascr_{\bm q}}}
$$
where $\bm\Ascr(\br)\equiv U(\br)\bm\nabla \Ud(\br)=-(\bm\nabla
U)\Ud$. Neglecting higher-order terms in $\bm q$, one has
$\G_{0,\bm p+\bm q}\mapsto\G_{0,\bm p}$; substituting $\sum_{\bm
p}\mapsto\nu\int \dd\xi$, one reduces the above expression to
\begin{equation}\label{der}
    \frac{\nu\vf^2\tel}{2d}\sum_{\bm q}\int\frac{(\xi
    \!+\! i\Lambda)\bm \Ascr_{\bm q} \cdot(\xi
   \!+\! i\Lambda)\bm \Ascr_{\bm q}}
    {\Lr(){\xi^2+1}2} \dd \xi = \frac{\pi\nu D}8\sum_{\bm
    q}\tr\Lr[]{\bm\Ascr_{\bm q},\,\Lambda}2=\frac{\pi\nu D}8
     \Tr\Lr(){\nabla Q}2,
\end{equation}
where $D=\vf^2\tel/d$  is the diffusion coefficient, and the
identity $\nabla Q=\Ud\lr[]{\bm\Ascr ,\,\Lambda}U$ was used, that
follows from the representation \Ref1 and the definition of
$\bm\Ascr$. This completes the derivation of the \N\ functional
\Ref{nlsm}, the matrix field $Q$ obeying constraints
\Ref{constraint} or, equivalently, \Ref 1. Including the
symmetry-breaking factors, like a magnetic field, is
straightforward  as I will demonstrate below when deriving the \N\
for interacting electrons. The source fields $\bm\eta$ were left
out of the expansion, but including them is simple: as seen from
\Eq s, they enter on par with the term $\omega\Lambda$, so that
their inclusion would result in additional gradientless terms,
$\Tr\bm \eta Q$. In a similar way, one can include any other
source fields, e.g.\ those necessary for direct evaluation of the
conductance \cite{AKL:86}.

The expansion \Ref{gradexp} was made to the first non-vanishing
powers in the small parameters \Ref{applicability}. The higher
order terms, both in gradients and in the (symmetry-breaking)
frequency may also be included. They are required in order to
obtain tails of different distribution function  or to describe
the long-time relaxation in the framework of the replica \N\
\cite{AKL:86} but this goes far beyond present considerations.

What is left to do, before turning to the interaction case, is to
illustrate how the \N\ works. The model is nonlinear because of
the presence of the constraint \Ref{constraint} -- so such an
illustration should show how the constraint may be explicitly
resolved. From all the examples listed in the introduction, I will
consider only one -- that of mesoscopic fluctuations
\cite{Al:85,L+S:85,AKL:86}.

\section{Mesoscopic fluctuations within the \N}
\label{diagrammatics}

The most famous of sample-to-sample (mesoscopic) fluctuations are,
of course, the universal conductance fluctuations
\cite{Al:85,L+S:85} whose full distribution was first calculated
within the replica \N\ \cite{AKL:86}. For the illustrative
purposes of this section, it is more convenient to consider
mesoscopic fluctuations of density of states (DoS), as this is the
simplest quantity to calculate, whether diagrammatically
\cite{AlSh:86} or within the \N\ \cite{AKL:86}. The main
simplification within the \N\ is that one does not need to
introduce any source field, since all the moments of the DoS
fluctuations, e.g. variance, can be obtained  by differentiating
the partition function \Ref z with respect to frequency $\omega$:
\begin{equation}\label{DeltaNu}
 \frac{\av{(\delta \nu)^2}}{\nu_0^2} =-\frac12\left.\lim_{n\to
0}\Fra (){\delta_0}{ \pi   n
}2\frac{\partial^2\lr<>{Z(\omega)}}{\partial\omega
^2}\right|_{\omega=0}\,.
\end{equation}
Here $\delta\nu\equiv\nu-\nu_0$,  $\nu_0$ is the
disorder-independent average DoS, and $\delta_0\equiv1/\nu_0L^d$
is the mean level spacing. To get \Eq{DeltaNu}, one first
expresses $\nu$ in terms of Green's functions,
$$
\nu(\varepsilon)=\frac{i}{2\pi L^d}\int\Bigl[{ G}^+_
\varepsilon(\,\b r,\b r) -{ G}^- _ \varepsilon(\,\b r,\b r)\,,
\Bigr]\,\dd^dr
$$
and using the representation of $G^\pm$ in terms of functional
integrals, \EQ{ZF}S one represents the above expression as
$$
\nu=\frac1{2\pi L^d}\,\frac1{Z_1}
\bpsiint\ee^{iS_1(\bm\eta,\omega)} \Tr\overline{\bm \psi}(\bm r )
\Lambda {\bm \psi}(\bm  r)=-\frac{i}{\pi L^d}\,\frac{\partial
Z_1(\omega)}{\partial\omega}\,.
$$
The expression for $\av{(\delta\nu)^2}$ is then obtained by the
replication and the ensemble averaging of the product of two DoS,
\Eq K, as described above. Then $Z(\omega) $ is finally expressed
as the partition function for the \N, \Eq z. It is implied in \Eq
{DeltaNu} that one considers the fluctuations in an open sample,
where an inevitable infrared divergence is cut off at $\omega\sim
E_{\text{T}}$ (which allows one to put $\omega\to0$) where the
Thouless energy $E_{\text{T}}=D/L^2$ is inversely proportional to
the diffusion time through the sample. For a closed sample,
\Eq{DeltaNu}  at finite $\omega$ can be used for calculating the
correlation function $\av{\nu(\varepsilon +\omega)\nu(\varepsilon
)}$ but for $\omega\lesssim\delta_0 $ considerations should be
more subtle (see \cite{KamMez:99a,YL:99a}) than what follows.

The simplest task is to calculate $Z(\omega)$ perturbatively. In
order to make any calculation, one has to resolve the constraints
\Ref{constraint}, i.e.\ to express $Q$ in terms of unconstrained
matrices. For the purpose of constructing regular perturbation
series, it is most convenient to parameterise $Q$ as follows
\cite{AKL:86}:
\begin{equation}\label{P}
Q=(\openone-W/2)\Lambda (\openone-W/2)^{-1}\,.
\end{equation}
 Here $W$
is anti-Hermitian matrix, off-diagonal in the
``retarded-advanced'' space, labelled by $+$ or $-$ indices as
described between \EQS n{SR}:
\begin{equation}\label{W}
W =  \begin{pmatrix}
 0&W^{+-} \\
W^{-+}&0
\end{pmatrix}\equiv \begin{pmatrix}
 0&B \\
 -B^+&0
\end{pmatrix}
\end{equation}
Here the $2n\times 2n$ matrices $B$ are unconstrained but for the
fact that they obey the charge conjugation symmetry,
\Eq{conjugate}. This means that they are ``real quaternion''
matrices, as described after \Eq 1:
$B=b^0\openone_2+ib^\mu\tau_\mu$, with $\tau_\mu$ being Pauli's
matrices in the charge conjugate space  and all $b^\mu$
($\mu=0,1,2,3$) being totally unconstrained $n\times n$ real
matrices. The parametrisation \Ref P makes $Q$ automatically
obeying both constraints in \Eq{constraint}, $Q^2=\openone$ and
$\tr Q=0$. The structure \Ref W makes $W$ anticommuting with
$\Lambda$, thus ensuring that $Q$ is Hermitian. The functional
integration in \Eq z should now be performed over independent
matrix components of $B$.  A particular convenience of the
parametrisation \Ref P is in the fact that in the $n=0$ replica
limit, the Jacobian of the change from $Q(\b r)$ to $B(\b r )\;$
is $1$. To perform the integration, one should (again!) reduce it
to Gaussian, by representing $\cF=\cF_0+\lr(){\cF-\cF_0}$. Here
the quadratic in $W$ functional ${\cal F}_0$ is obtained by
expanding the \N\ functional \Ref{nlsm}  to the lowest order in
$W$:
\begin{equation}\label{B}
{\cal F}_0=- \frac{\pi\nu_0 D}8\int\tr(\,\nabla\! W)^2\,\dd^dr=
\frac{\pi\nu_0 D}4\Tr|\nabla\!B|^2\, .
\end{equation}
The $\omega$-proportional term is not included into $\cF_0$
(although it should, if one were interested in frequency-dependent
observables),  since in \Eq{DeltaNu} one needs only quadratic in
$\omega$ terms.  They are obtained by expanding
$\ee^{-\lr(){\cF-\cF_0}}$ in \Eq z for $Z(\omega)$ in powers of
$\lr(){\cF-\cF_0}$ and further expanding
$Q=\Lambda(1+W+\frac12W^2+\ldots$ according to \Eq W, and finally
performing the functional integration
\begin{equation}\label{a}
\left<\ldots\right>_0\equiv\frac{\int{\cal D}B(\ldots)\exp
(-\cF_0)} {\int{\cal D}B\exp (-\cF_0)}\,.
\end{equation}
As this integration is Gaussian, the result would be  a product of
all pair averages.  The later can be found immediately, as \Eq B
ensures that the the functional average \Ref a splits into a
product of Gaussian integrals over independent matrix elements
$b^\mu_{ij}$ of $B$:
\begin{equation}\label{b}
   \av{b^\mu_{ij}(\b r)b^\nu_{km}(\b r')}_0=\Pi(\b
r -\b r^\prime)\delta^{\mu\nu}\delta_{ik}\delta_{jm}\,,
\end{equation}
where the propagator $\Pi(\b r\E- \b r')$ obeys the diffusion
equation, \mbox{$\pi\nu_0D\nabla^2 \Pi(\b r\E- \b r')=\delta(\b r
\E- \b r')$}, with proper boundary conditions (open sample). For
an $L^d$ cube,  $\Pi$ can be defined via its Fourier transform as
\begin{equation}\label{D}
\Pi(\b r -\b r^\prime)=\frac{\delta_0}\pi \sum_q\frac{1}{
{D}q^2}\,\ee^{ i\b q\cdot (\b r-\b r^\prime)} \,.
\end{equation}
\Eq b is relatively easy to derive but very inconvenient to use,
since all matrices that enter the averaging should be explicitly
split into their components in the retarded-advanced space, like
in \Eq W, and in the charge conjugate space, like in \Eq {dc}. But
all these components can be assembled together once and forever by
casting the averaging formula \Ref b into those (two) written
explicitly in the full matrix space of $Q$ \cite{AKL:86}:
 \begin{gather}
\Bigl<\tr \Bigl(W_{\b r}\,P\, W_{\b r'}\,R\Bigr)\Bigr>_0 = \Pi(\b
r -\b r^\prime) \Bigl\{\,\tr (\Lambda P\Lambda R^+-PR^+) +\tr
(\Lambda P)\tr (\Lambda R)-
\tr (P)\tr (R)\Bigr\} \nonumber  \\
\Bigl<\tr \Bigl(W_{\b r}\,P\Bigr)\,\tr\Bigl(  W_{\b
r'}\,R\Bigr)\Bigr>_0= \Pi(\b r -\b r^\prime) \tr (\Lambda P\Lambda
R-\Lambda P\Lambda R^+ -PR+PR^+) \label{V}
\end{gather}
Here $P$ and $R$ are arbitrary  matrices of the same rank as $W$
(or $Q$). Equations \Ref V allow one to calculate any perturbative
(or RG) expression. Let us use it for calculating $Z(\omega)$ up
to $\omega^2$ and thus finding the DoS variance \Ref{DeltaNu}.

\begin{figure}[t]
  \begin{center}{
  \scalebox{.74}{\includegraphics*[clip,angle=0]{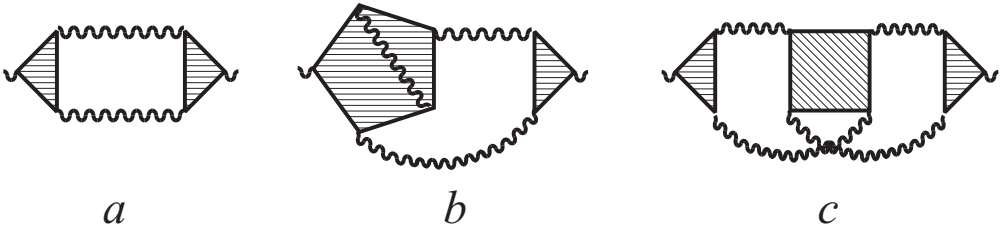}}}
   \end{center}

\caption{One-loop ($a$) and two-loop ($b$ and $c$) contributions
to the DoS variance: hatched polygons with $2n+1$ vertices
correspond to the expansion of $\tr ({\omega} {\Lambda} Q)$ to the
$2n^{\text{th}}$ power in $W$, with external vertices representing
the source field ${\omega} $; hatched $2n$ polygons correspond to
the appropriate expansion of $\tr( \nabla Q)^2$; the wavy lines
represent $W$ pairings, i.e.\ the diffuson propagators \Ref{D}, in
correspondence with the standard
 diagrammatic techniques, Fig.~\ref{diffuson}.}
\label{dos} \end{figure}

In the lowest perturbative order one needs to calculate the
functional average $\frac12\av{\cF_{\omega}^2}_0$, where
$\cF_{\omega}$ is the $2^{\text{nd}}$ term in \Eq{nlsm}. Expanding
$Q$, one finds that
\begin{equation}\label{O}
\cF_{\omega}\equiv\frac{i\pi\nu_0\omega }4\Tr\Lambda
Q=\frac{i\pi\nu_0  }8\lr[]{\Tr\lr(){\omega W^2} +\frac12
\Tr\lr(){\omega W^4} +\ldots}
\end{equation}
  since the zeroth term is
an irrelevant $n$-proportional constant, and all the odd order
terms vanish as $\tr W=0$. It is convenient to put ${\omega} $
inside the $\Tr$: in general, source fields that are necessary to
calculate any other quantity but DoS are matrix field, and in a
similar expansion they would have been inside the trace. Now to
the lowest order
\begin{equation}
\label{L} Z(\omega) = -\frac{ \pi^2\nu_0 ^2 }{128}
\av{\Tr\lr(){\omega W^2}\, \Tr\lr(){\omega W^2} }_0\,.
\end{equation}
One applies the averaging formulae \Ref{V} by choosing all
possible pairings and performing them one after another (in an
arbitrary order), considering at each step all matrices $W$ (but
two active, i.e.\ participating in the chosen pairing) as fixed.
There are only two different pairing schemes possible in \Eq L:
one either make two inter-trace, or two intra-trace pairings. In
the first case, one applies first of the formulae \Ref{V}, with
$P=R=\openone$; then only the last term does not vanish and equals
$(\tr 1)^2=16 n^2$. Making two such inter-trace averaging one
finds the contribution proportional to $n^4$ which vanishes in \Eq
{DeltaNu} in the replica limit. Therefore, one needs only the
intra-trace averages which is convenient to represent
diagrammatically, Fig.~\ref{dos} ($a$). In such a representation,
each trace corresponds to a (hatched) polygon with vertices
representing either matrices $W$ or the source field $\omega$ (no
more than one vertex in each polygon). At the first step of
averaging, one applies the second of \Eq V with $P\equiv \omega
W(\b r)$ and $R\equiv \omega W(\b r')$. Taking into account a
combinatorial factor of $2$ (each $W$ in the first trace can be
paired with either of two in the second trace) and using that $W$
is anti-Hermitian and commutes with $\Lambda$, one finds
\begin{multline*}
    \av{\Tr\omega W^2 \Tr\omega W^2}_0=-8\omega^2\ir{\Pi_{\b r-\b r'}\av{
    \tr W_{\b r}  W_{\b r'}}_0}\dd^d r'=8\omega^2\Lr(){\tr 1}2
    \ir{\Pi^2_{\b r-\b r'}}\dd^d r' \\
   =128 \omega^2n^2 L^d\ir{\Pi^2_{\b r}}=\frac{128 \omega^2n^2}{\pi^2\nu_0^2}
   \sum_{\b q}\frac1{\Lr(){Dq^2}2}\,.
\end{multline*}
Combining this with \Eq L and substituting the result into \Eq
{DeltaNu}, one recovers the well-known result \cite{AlSh:86}:
\begin{equation}\label{nn}
    \frac{\av{(\delta \nu)^2}}{\nu_0^2} =
\Fra(){\delta_0}\pi2\sum_{\b q}\frac1{\Lr(){Dq^2}2}=
\frac{C_d}{g^2}\,,
\end{equation}
where $g=2\pi^2\nu_0D L^{d-2}$ is the dimensionless conductance of
a cube of volume $L^d$, and $C_d$ is a numerical coefficient
dependent on the sample shape and dimensionality.

Exactly the same technique could be applied for calculating the
two-level correlation function, $\av{\nu(\varepsilon
)\nu(\varepsilon +\omega)}/\nu_0^2 -1$. In this case $\omega$
plays a dual role of the frequency and the source field. The only
technical change is that the $\Tr \omega W^2$ should be included
into the zeroth approximation, \Eq B, which results in replacing
$Dq^2$ by $Dq^2-i\omega$ in the denominator of \Eq D. In the
diffusive regime of relatively large $\omega$, i.e.\
$\omega\gtrsim D/L^2$, the result \Ref{nn} remains roughly valid,
with substituting $L_\omega\equiv(D/\omega)^{1/2} $ for $L$ into
the expression for $g$. However, this would fail in the
two-dimensional case where for $L_\omega\ll L$ the coefficient
$C_2$ in \Eq n vanishes due to the analytic properties of the
diffusion propagator $\Pi(\omega, \b q)$ \cite{KL:95a}. In this
case, the main perturbative contribution is due to the two-loop
diagrams, Fig.~\ref{dos} ($b,\, c$). The hatched pentagon in
diagram ($b$) corresponds to the $2^{{\text{nd}}}$ term in the
expansion \Ref{O}. The hatched square in diagram ($c$) results
from the average $\big<\cF_\omega^2\cF_\nabla^{(4)}\big>_0$ with
$\cF_\nabla^{(4)}$  obtained by expanding $\nabla Q^2$ to the
$4^{\text{th}}$ power of $W$. Such expansions give all the
necessary coefficients corresponding to the Hikami boxes of the
standard diagrammatic technique,  so that one is left with
calculating two-loop integrals containing $3$ or $4$ diffusion
propagators. Note that in the diagrammatic technique the diagrams
$b$ and $c$ cancel each other while the only contribution results
from the third possible 2-loop diagram \cite{KL:95a} describing
the weak localisation correction to one of the diffusons. Such a
diagram, on the contrary, vanishes within the parametrisation
\Ref{P} while the identical result is given by the remaining
two-loop diagrams $(b, \, c)$.

Finally note that the RG calculations are also based on the
averaging formulae \Ref{V} which could also be appropriately
modified to include cases of partially broken symmetry, either due
to the presence of a magnetic field or magnetic impurities
(unitary symmetry) or to the presence of spin-orbit scattering
(symplectic symmetry). In all cases, the same averaging formulae
are also applicable to the case when the \N\ also encompasses
interactions.

\section{Coulomb and pairing interactions in the sigma model}
  \label{interactions}

  The \N\
works perfectly for perturbative problems and -- in certain cases
-- does reasonably well even for  non-perturbative problems within
the model of non-interacting electrons in the presence of
disorder. Since the inclusion of interactions beyond the
mean-field approach is impossible within the supersymmetric
method, the replica approach is still the most reliable tool for
models with interactions.

The original fermionic replica \SM\ \cite{EfLKh} has been
generalised by Finkelstein \cite{Fin} to include the interactions.
The interest in this approach has been greatly enhanced by the
recent discovery \cite{Krav:95} of an apparent metal-insulator
transition in 2D disordered systems in zero magnetic field. It is
not at all clear whether the observed effects are, indeed, due the
transition \cite{Sav+Pep2,Pep:1999}, and if they are -- whether
such a transition is, indeed, driven by interactions. However, a
possibility of having the metal-insulator transition in a 2D
disordered interacting system (schematically described by the TOE
model \Ref{TOE}), while it is absent in a disordered
noninteracting system where all states are localised, is
intriguing by itself. Such a possibility is  arguably  a driving
force in a considerable revival of interest in Finkelstein's
$\sigma$ model (see, e.g.,
Refs.~
\onlinecite{CdCKL:98,AnKam:99,Lud:98,FLS:00,YL:01}).
At the moment, there is no clear evidence whether Keldysh
techniques employed in  Refs.~\cite{AnKam:99,Lud:98} would give
any edge over the replica method described in these lectures. So
all further considerations are also limited to the replica
approach.

Below I  describe how to include the BCS pairing interaction and
the Coulomb interaction (for simplicity, in a singlet channel
only, as the inclusion of a triplet channel is technically almost
identical) into the derivation of \N\ given in section \ref{NF}.
One inevitable complication is that such a derivation should be
done in the temperature technique, and here some leap of faith is
necessary in order to come to any result within a relatively short
paper. The resulting \SM\ looks similar to that for the
non-interacting model, although its usage is considerably mode
difficult. Still, I give a relatively simple example of such a
usage to derive some known result for the BCS model. Note that the
derivation follows that in Ref.~\cite{YL:01} which is similar in
spirit to that in Ref.~\cite{AnKam:99} but rather different from
the original derivation in Ref~\cite{Fin}. The derivation below is
also made as parallel as possible to that of the \N\ in the
absence of interactions.

\subsection{Hubbard--Stratonovich decoupling}
Since the pair interaction, in contrast to the elastic scattering
from impurities, does not conserve single-particle energy, the
effective interaction functional should be dynamical. As usual, it
is convenient to introduce imaginary time $\tau$ implying
everywhere the thermodynamic Gibbs averaging (together with the
averaging over quenched disorder where applicable). Thus one
considers thermodynamic Green's functions related (in the
Matsubara frequency representation) to the retarded and advanced
Green's functions of the previous sections by the standard
procedure of analytical continuation \cite{AGD}. Then the
effective functional corresponding to the interaction term in
\Eq{TOE} can be written as
\begin{equation}\label{EfFunc}
  \cF_{\text{int}}=\frac12\Int\dd x
  \dd
  x'\bar\psi_s(x)\bar\psi_{s'}(x')V_{xx'}\psi_{s'}(x')\psi_s(x)\,,
\end{equation}
where $x\equiv \{\b r, \tau\}$, and all the fermionic fields are
anti-periodic in imaginary time $\tau$ with period $1/T$,
$s=(\uparrow,\,\downarrow)$ is the spin index,
$V_{xx'}\equiv\delta(\tau-\tau')V_c(\b r-\b r')$, and $V_c$
represents the Coulomb interaction. The fields are still assumed
to have the same matrix structure, \Eq{conjugate}, as in the
noninteracting case, necessary to take into account both the slow
modes (diffuson and cooperon) arising from the impurity
scattering. After the replication necessary for treating the
random potential, all the fields in \Eq{EfFunc} naturally have the
same replica index and the action would be just a sum of all the
actions \Ref{EfFunc} with different replica indices. Similarly,
the BCS functional describing the attraction in the Cooper channel
can be written as
\begin{equation}\label{BCSFunc}
\cF_{\text{sc}} =\lambda_0\int\!{\rm d}x\, \bar\psi_{\uparrow} (x)
\bar \psi_{\downarrow} (x) \psi_{\downarrow}(x)\psi_{\uparrow}(x),
\end{equation}
where $\lambda_0$ is the BCS coupling constant. The total action
is now the sum of

The Hubbard--Stratonovich decoupling of the functional
\Ref{EfFunc}-\Ref{BCSFunc} is similar to that for the disorder
functional \Ref{g}, i.e.\ some new matrix fields are introduced to
decouple the quartic terms via the Gaussian integration like in
\Eq g. To perform this simultaneously with averaging over
disorder, one needs to replicate all the fermionic fields as
described after \Eq n. All the quartic interaction terms in the
action are naturally diagonal in the replica indices.

To allow for all the slow modes in the interaction functional
\Ref{EfFunc} one should introduce three matrix fields: the fields
$\Phi$ and $\hat f$ to take account of a small-angle scattering (a
singlet channel) and large-angle scattering, and yet another one
that corresponds to the Coulomb repulsion in the Cooper channel.
The last one would lead to the standard renormalization of the BCS
attraction (see, e.g., Ref.~\cite{Nagaosa}) and is not considered
here; further, it is assumed that systems under considerations
still have an effective attraction in the Cooper channel after
such a renormalization. The interaction in the Cooper channel is
then decoupled with the help of another matrix field, $\Delta$.
Then the decoupling has the following form:
\begin{align}
\ee^{-\cF_{\text{int}}}&=\Int\D\Phi\exp\left\{\!-\frac{1}{2}\!\Int\dd
x\dd x'\, \Phi(x) V^{-1}_{xx'}\Phi(x') +i\!\Int\dd
x\,\Psib_{s}(x)\Phi(x)\Psi_{s}(x)\right\} \nonumber\\&+ \Int\D\!
\hat f\exp\left\{\!-\frac{1}{2}\!\Int\dd x\dd x'\,\tr\!\lr[]{\hat
f(x) V^{-1}_{xx'}\hat f(x')} +i\Int\dd
x\,\Psib_{s}(x)f_{ss'}(x)\Psi_{s'}(x)\right\}\!,
\label{Sfields}\\\nonumber
\ee^{-\cF_{sc}}&=\Int\D\Delta\exp\biggl\{-\frac{1}{\lambda}\int\dd
x \left|\Delta(x)\right|^2 +i\Int\dd
x\left[\Delta(x)\Psib_{\uparrow}(x)\Psib_{\downarrow}(x)
-\bar\Delta(x)\Psi_{\downarrow}(x)\Psi_{\uparrow}(x)\right]\biggr\}
\end{align}
After performing the Gaussian integration here, together with that
in \Eq g, one obtains instead of the action \Ref{Trln} the
following effective action that depends both on $\sigma$ (which is
a new name for $Q$, the latter being reserved for a later usage)
and on all the new fields:
\begin{equation}
\label{Str} \widetilde
\cF=\cF_{\text{fields}}+\frac{\pi\nu}{8\tau}\Tr \sigma^2 -
\frac{1}{2}\Tr\ln\left[ \hat{\xi} -i\left(\frac{1}{2\tau}\sigma
 + \Phi + \Delta +\hat\varepsilon\right)- \hat f\right]\,.
\end{equation}
Here $\cF_{\text{fields}}$ is the quadratic in $\Phi$, $f$ or
$\Delta$ part of the action \Ref{Sfields}; the operator $\hat
\varepsilon$ equals $i \hat \tau_3
\partial _\tau$ in the imaginary time representation with $\tau_3$ being
the appropriate Pauli matrix in the charge-conjugate space defined
by \Eq{conjugate} (it becomes a diagonal matrix of fermionic
Matsubara frequencies, $\varepsilon_n=\pi (2n+1)T$, in the
frequency representation); $\Tr $ refers to a summation over all
the matrix indices and to an integration over $\bbox r$ and $\tau$
(or summation over Matsubara frequencies in the frequency
representation).  The triplet channel, described by the field
$\hat f$, is quite important: in particular, it can lead to the
delocalization in the presence of disorder \cite{Fin,CdCKL:98}.
Nevertheless,, such effects will not be considered here and this
term will be ignored from now on -- mainly for simplicity (also,
it does not contribute when considering effects of the Coulomb
interaction in superconducting systems). The field $\sigma$ has
the same structure as $Q$ in \EQS{dc}{1}, apart from explicitly
including the $2\times2$ spin sector and replacing the $2\times2$
retarded-advanced sector by the dependence on the imaginary time
$\tau$ (since it is diagonal in $\tau$, it becomes a matrix field
in the Matsubara frequencies). In the absence of the fields $\Phi$
and $\Delta$, the saddle-point solution for the action \Ref{Str}
is formally the same as in the zero-temperature techniques of
section~\ref{NF}, \Eq 1,  but the matrix $\Lambda$ has a non-unit
structure in the Matsubara (instead of advanced-retarded) sector:
\begin{equation}\label{SP2}
 \sigma=U^\dagger\Lambda U\,,\qquad
\Lambda ={\text {diag\,\{sgn}}\,\hat\varepsilon\}\,.
\end{equation}
The field $\Phi$, corresponding to the singlet part of the Coulomb
interaction, is diagonal in all the sectors. The
``order-parameter'' field ${\hat \Delta}$ is   Hermitian and
self-charge-conjugate, diagonal in the replica indices and
coordinates $\mathbf r$ and $\tau$, and has the following
structure in the spin and time-reversal space:
\begin{equation}
{\hat \Delta}(x)= |\Delta(x)| e^{\frac{i}{2}\chi(x)\hat\tau_3}
\,\hat \tau^{{\text{sp}}}_2\otimes\hat\tau_2^{\text{sp}}\,
e^{-\frac{i}{2}\chi(x)\hat\tau_3}\,, \label{Delta}
\end{equation}
where $|\Delta |$ and $\chi$ are the amplitude and the phase of
the pairing field $\Delta(\b r, \tau)$, $\hat\tau _i$ and
$\hat\tau^{\text{sp}}_i$ are Pauli matrices ($i=0,1,2,3$ with
$\hat\tau^{}_0=1$) that span the charge-conjugate and spin
sectors, respectively.

When $\Phi$ and $\Delta$ are included, the saddle-point equation
for the functional \Ref{Str} can be formally written similarly to
\Eq{SP} for the non-interacting zero-temperature case:
\begin{equation}
\label{sp} -i\pi\nu\sigma({\mathbf r}) = \left\langle{\bm r}\left|
\left[{-{\hat{\xi}}+ \frac{i}{2 \tau_{\text{el}}}\,\sigma+
i\left({
 \hat{\varepsilon}} + {\hat \Delta}+\Phi\right)}
\right]^{-1}\right|{\bm r}\right\rangle
\end{equation}
Now $\sigma=\Lambda$  does no longer represent the saddle point
solution for $\varepsilon\ne0$. Still, one can derive an effective
functional by expanding the above Tr ln within the manifold
\Ref{SP2} in the symmetry-breaking field $\epsilon+\Delta+\Phi$
and in gradients of $\sigma$, as has been done in the original works by
Finkelstein \cite{Fin}. An alternative is to make first a
similarity transformation around $\Lambda$ within the manifold
\Ref{SP2} to find the saddle point solution $\sigma_{\text{sp}}$
in the presence of the fields. This can be formally done with the
help of matrix $U_0$ that diagonalizes the Hermitian field
$\hat\varepsilon + \hat \Delta + \Phi$:
\begin{equation}\label{diag}
\hat\varepsilon + \hat \Delta + \Phi=U_0^\dagger \lambda
U_0\,,\qquad \sigma_{\text{sp}}=U_0^\dagger \Lambda U_0\,.
\end{equation}
Here $U_0$ (which is yet undetermined but fixed) belongs to the
same symmetry group that defines the manifold \Ref{1} in the
absence of the interaction fields. By substituting \Eq{diag} into
the saddle-point equation \Ref{sp}, one can easily verify that
this is, indeed, a spatially-homogeneous solution, provided that
$\lambda \tel\E\gg1$ which will always be the case for a dirty
superconductor ($\Delta\tel\E\gg1$) or a dirty metal
(dimensionless conductance $g\E\gg1$).

Now one can perform the expansion of Tr ln, \Eq{Str}, in gradients
of $Q$ and in the symmetry-breaking fields represented by the
eigenvalues $\lambda$ and the matrix $U_0$, \Eq{diag}, by
employing the following parametrisation:
\begin{equation}
\label{sig} \sigma= U_0^\dagger Q^{\,}U^{\,}_0\,, \qquad Q=
U^\dagger  \Lambda U\,,
\end{equation}
where $Q$ represents the saddle-point manifold in the metallic
phase and $\sigma$ is obtained from $Q$ by the same rotation
(\ref{diag}) as $\sigma_{\text{sp}}$ is obtained from the metallic
saddle point $\Lambda$. Therefore, $Q$ is defined, as in the
metallic phase, on the coset space $\S(2n)/\S(n)\otimes \S(n)$
where, depending on the symmetry, $\S$ represents the unitary,
orthogonal or symplectic group. In the noninteracting case of
section~\ref{NF}, only the case of the symplectic group,
corresponding (unfortunately) to the orthogonal symmetry, was
described, but a generalisation to the other two cases is
straightforward and described in almost any paper on the \N\ for
disordered systems. The parameterisation (\ref{sig}) simplifies
considerably all the subsequent derivations and leads to a new
variant of the nonlinear $\sigma$ model for interacting systems
\cite{YL:01} which can be more convenient for many applications
than the original one \cite{Fin}. After substituting
$\sigma=U_0^\dagger U^{\dagger}_{\phantom{0}\,}
\Lambda^{\,}U^{\,}U^{\,}_0$, \Eq{sig}, into \Eq{Str}, one obtains
the following representation for the Tr$\,$ln term:
$$
\delta {\cF} = -\tfrac 12 {\mathrm Tr}\ln \bigl\{\!\hat G_0^{-1}
+U^{\,}_ 0[\hat\xi,U^\dagger_0]-i(U\lambda U^\dagger) \bigr\}\,,
$$
where one can also include an external magnetic field with the
vector potential $\b A$:
$$
\hat G_0\equiv \left( \hat \xi -\frac{
i}{2\tau_{\text{el}}}\Lambda \right)^{-1},\qquad
\hat{\xi}\equiv\frac1{2m}(\b p -e\b A)^2-\ef\,.
$$
The expansion to the lowest powers of gradients and $\lambda$ is
now straightforward and similar to that for the noninteracting
case. It results after some calculations in the following
effective action:
\begin{eqnarray}
\widetilde{\cF}= \frac{1}{2}\Tr\!\lr[]{ \Phi V^{-1}_{\b r\b r
'}\Phi}+\frac{1}{T\lambda_0}\sum_{\omega}\int\!\!{\mathrm d}{\b
r}\,
|\Delta_{\omega}|^2 
+\frac{\pi\nu }{2} {\mathrm Tr}\!\left[\frac D4\left(\partial
Q\right)^2 - \lambda Q\right]\,. \label{new}
\end{eqnarray}
The long derivative in Eq.~(\ref{new}) is defined as
\begin{equation}
\partial Q
\equiv \nabla Q + \Bigl[{\b  A}_0 \!\!-\! {ie} {\b  A}\hat\tau_3,
\,Q \Bigr] \equiv \partial _0 Q + \left[{\b  A}_0 , Q \right] \,,
\label{LD}
\end{equation}
where the matrix ${\b  A}_0$ is given by
\begin{equation}
\label{A} {\b  A}_0 = U^{\,}_0\partial^{\,}_0 U^\dagger_0 \,,
\end{equation}
and $\partial_0\equiv\nabla-[ie\b A \hat\tau_3,\ldots] $ is the
long derivative (\ref{LD}) in the absence of the pairing field
$\Delta$. Both $U^{\,}_0$ and $\lambda$ should be found from the
diagonalization of $\epsilon + \Delta+\Phi$, Eq.~(\ref{diag}).
Although such a diagonalization cannot be done in general, it will
be straightforward in many important limiting cases. For
$\Delta=\Phi=0$, the field ${\b  A}_0$ vanishes,
$\partial\to\partial_0$, and $\lambda\to\varepsilon$, so that the
functional (\ref{new}) goes over to that of the standard nonlinear
$\sigma$ model for non-interacting electrons.

The action \Ref{new} is most general in the current context. It
allows one to develop a fully self-consistent approach to
superconductivity of dirty metals in the presence of Coulomb
interaction. However, any application requires a set of further
simplifications. As a simple illustration, I will show below how
to use the model in a dirty superconductor near the
metal-superconductor transition in the absence of Coulomb
interaction.

\subsection{Ginzburg-Landau Functional}

In the vicinity of the metal-superconductor transition one can
expand the action (\ref{new}) (in the absence of the Coulomb field
$\Phi$) in the pairing field $\Delta$. A further simplification is
possible in the weak disorder limit, $g\gg1$, when   the field $Q$
can be integrated out to obtain an effective action for the
$\Delta$-field only. In the quadratic in $\Delta$ approximation,
the kernel of this action will give, with due account for the
disorder, an effective matrix propagator of the pairing field
which governs properties of a disordered superconducting sample
near the transition.

To integrate over $Q$, one splits the action (\ref{new})
 into $\widetilde{\cF} \equiv {\cF} + {\cF} _\Delta$ where
$\cF $ is the standard \N\ functional as in the metallic phase.
Then one makes a cumulant expansion, i.e.\ first expands ${\rm
e}^{-({\cF} + {\cF} _\Delta)}$ in powers of ${\cF} _\Delta$, then
performs the functional averaging with ${\rm e}^{- {\cF} _0}$
(denoted below by $\langle\ldots\rangle_Q$) and finally
re-exponentiates the results. Apart from the last step, this is
exactly as described in section \ref{diagrammatics} for the
noninteracting \N. The expansion involves only the first and
second order cumulants since the higher order cumulants generate
terms of higher order in $\Delta$. Then the only contributions to
the action quadratic in $\Delta$ are
\begin{eqnarray}
\lefteqn{\cF_{\rm eff}[\Delta] =\frac{1}{\lambda_0
T}\sum_{\omega}\int \!\!{\mathrm d}{\b  r}\, |\Delta_{\omega}|^2
-\frac{\pi\nu}{2}\Bigl\langle{\mathrm Tr}\, (\lambda
\!-\!\epsilon) Q\Bigr\rangle_{\!Q} }
\nonumber
\\[-3mm]
\label{av}
\\[-3mm]
\nonumber
&&-\biggl\langle\frac{\pi\nu D}{8}{\mathrm Tr}\,\left[{\b
A}_0,Q\right]^2 +\frac{(\pi\nu D)^2}{8}\Bigl( {\mathrm Tr}\,
Q\partial _0 Q {\b  A}_0 \Bigr)^{\!2}\biggr\rangle_{\!\!Q}.
\end{eqnarray}
Expanding $\lambda$ and ${\b  A}_0$ to the lowest power in $\Delta
$ and performing the functional averaging  one finds \cite{YL:01}
the action quadratic in $\Delta$ as follows:
\begin{equation}
\cF_{ {\text{eff}}}[\Delta] =\frac{\nu}{T}\sum_{\omega}\int\!{\rm
d} {\b r}\, \Delta^*_{\omega}({\b  r})\bigl<{\b
r}\bigl|\,\hat{\cal K}_{\omega} \,\bigr|{\b  r}'\bigr>
\Delta_{\omega}({\b  r}')\,, \label{K2}
\end{equation}
with the operator $\hat{\cal K}_{\omega}$ given by
\begin{equation}
\label{wl}
\hat{\cal K}_{\omega}=\frac{1}{\lambda_0 \nu} -2\pi T\!\!\!\!
\sum_{\epsilon(\omega-\epsilon)<0} \left\{ \hat\Pi^{c}_{\omega}
+\frac{1}{\pi\nu} \frac{ \Pi^{d}_{|2\epsilon-\omega|}(0){\hat
{\cal C}}}{(2\epsilon-\omega)^2} \right\}.
\end{equation}
Here $\Pi^{c,d}_{|\omega|}({\b  r}, {\b  r}') =\bigl<{\b
r}\bigl|\hat\Pi^{c,d}\bigr|{\b  r}'\bigr> $ are the cooperon and
diffuson propagators.  At $\omega=0$  the later coincides with
that of the \N\ for non-interacting electrons, \Eq D. In general,
\begin{equation}
\label{Pi}
\hat\Pi^{c}_{|\omega|} = \Bigl( {\hat{\cal C}} + |\omega|
\Bigr)^{-1}\,,
\end{equation}
where $ {\hat {\cal C}}\equiv -D\left(\nabla\E - {2ie}{\b
A}\right)^2$ defines the cooperon modes; $\hat\Pi^{d}$ is obtained
from $\hat\Pi^{c}$ by putting the external vector potential $\b
A=0$. In the last term in Eq.~(\ref{wl}), $\Pi^{d}_{|\omega|}
(0)\E\equiv \Pi^{d}_{|\omega|}({\b  r},{\b  r})$; this term may be
obtained by expanding in $g^{-1}$ the cooperon propagator with the
renormalised diffusion coefficient,
$$
{\hat{\cal C}} \to
\left[1-\frac{1}{\pi\nu}\Pi^{d}_{|\omega|}(0)\right]\,{\hat{\cal
C}},
$$
which is a weak localisation correction to free cooperon
propagator $\Pi^{c}_{|\omega|}({\b  r}, {\b  r}')$.

The summation over Matsubara frequencies in Eq.~(\ref{wl}) yields
\begin{equation}
\hat{\cal K}_{\omega}= \ln\frac{T}{T_0}+\psi\!\!\left(\!
\frac{1}{2}+\frac{|\omega|\!-\!{\hat{\cal C}} }{4\pi T}\! \right)
-\psi\!\!\left(\frac{1}{2}\right) - \frac{a_{\omega}{\hat{\cal
C}}}{4\pi T} \,, \label{K1}
\end{equation}
where $T_0\equiv T_{c0}(B\!=\!0)$ is the transition temperature of
the clean superconductor in the absence of a magnetic field and
$\psi$ is the digamma function. The weak localisation correction
is proportional to the coefficient $a_{\omega}$ given by
$$
\begin{array}{l}
\displaystyle a_{\omega}(T)=\frac{1}{\pi\nu
V}\sum_{\mathbf{q}}\frac{1}{Dq^2}\left\{
\psi'\left(\frac{1}{2}+\frac{|\omega|}{4\pi T}\right) \right. \\[6mm]
\displaystyle \left. -\frac{4\pi T}{Dq^2} \left[
\psi\left(\frac{1}{2}+\frac{|\omega|+Dq^2}{4\pi T}\right)
-\psi\left(\frac{1}{2}+\frac{|\omega|}{4\pi T}\right) \right]
\right\}.
\end{array}
$$
For $\omega=0$ the coefficient $a_0\equiv a_{\omega =0}(T)$ can be
simplified in the two limits:
\begin{equation}
 a_0=\left\{
\begin{array}{ll}
\displaystyle \quad\frac{\psi'({1/2})}{\pi\nu
L^d}\!\!\!\sum\limits_{L_{T}^{-1} <q<\ell^{-1}}\frac{1}{Dq^2}\,, &
L \gg L_T\,,\\[8mm]
\displaystyle -\frac{\psi''({1/2})}{8\pi^2\nu L^d T}\,\,, & L \ll
L_T \,,
\end{array}\right.
\end{equation}
where $L_T\equiv\sqrt{D/T}$ is the thermal smearing length.

The instability of the normal state (i.e. a transition into the
superconducting state) occurs when the lowest eigenvalue of the
operator ${\hat{\cal K}}_{\omega}$ becomes negative. The
eigenfunctions of this operator coincide with the eigenfunctions
of the cooperon operator ${\hat{\cal C}}$. The lowest eigenvalue
of ${\hat{\cal C}}$
 is known to be ${\cal C}_0= DB/\phi_0$, where
$\phi_0$ is the flux quanta. This ground state cooperon
eigenfunction corresponds to the lowest eigenvalue ${\cal K}_0$ of
the operator ${\hat{\cal K}}_{\omega}$. The condition ${\cal
K}_0=0$ implicitly defines the line $T_c(B)$ in the $(T,B)$-plane
where the transition occurs:
\begin{equation}
\ln\frac{T_c}{T_0} +\psi\left(\frac{1}{2}+\frac{{\cal C}_0}{4\pi
T_c}\right) -\psi\left(\frac{1}{2}\right)=\frac{a_0 {\cal
C}_0}{4\pi T_c}. \label{Tc(B)}
\end{equation}
The term in the r.h.s.\ of Eq.~(\ref{Tc(B)})
 describes a $1/g$-correction to the main result. This
 weak localisation is linear in the magnetic field $B$ and
vanishes as $B\to 0$ as expected (the Anderson theorem
\cite{And:59}). In a nonzero magnetic field the weak localisation
correction to $B_c$ is positive which has a very simple
explanation. The superconductivity is destroyed by the magnetic
field when $\Phi(\xi)\gtrsim\Phi_0$, where $\Phi(\xi)$ is the flux
over the area with linear size of the order of the coherence
length $\xi\sim\sqrt{D/T}$ and $\Phi_0$ is the flux quanta. The
weak localisation corrections reduce $D$ and thus  $\xi$.
Therefore, one needs a stronger field to destroy the quantum
coherence. The same reasoning explains the growth of $T_c$ in a
fixed magnetic field.

Note finally that it would be straightforward to include the
leading weak-localization corrections in all orders of $g^{-1}\ln$
by calculating the $Q$ - averages in \Eq{av} via the
renormalization group. This would lead to the renormalization of
$D$ in the cooperon propagator (\ref{Pi}), thus changing the shape
of the $T_c(B)$ curve. However, the value of $T_c(0)$ will again
remain unaffected, since the superconducting instability is
defined by the onset of the homogeneous zero mode in the operator
$\hat {\cal K}$, Eq.~(\ref{K1}), which does not depend on the
value of the diffusion coefficient in the cooperon propagator.

It is worth stressing that the Anderson theorem reflects certain
properties of a model rather than those of real superconductors.
If one allows for Coulomb interaction, then the critical
temperature of the superconducting transition is no longer
independent of disorder. Combined effects of the interaction and
disorder lead to corrections to the transition temperature
proportional (in a slightly simplified way) to $g^{-1}\ln
^3(T_c\tel)$ \cite{Ovchin:73,MEF:84,Fin:87}. For a relatively weak
disorder, the system still remains superconducting at $T\to0$
while for a sufficiently strong disorder the above corrections
would suppress the superconducting pairing at any temperature and
make a 2D system insulating at $T\to0$. Such a mechanism of
suppressing $\Delta$ by disorder gives a possible scenario for a
widely observed super\-conductor--insulator (SI) transition in
two-dimen\-sional structures [58--61] 
which is most adequately described within the \N\
\cite{Fin:87,Fin} similar to those developed in this section.
However, a widely accepted scenario for such a transition is based
on very different, both in origin and in implementation, models
\cite{AES:82,LO:83,AES:84,MPAF:86,FGG:90}
where $|\Delta|$ remains finite and the superconductivity is
suppressed by the loss of the phase coherence due to the
fluctuations of the phase $\chi$ of the order parameter. These
``phase-only'' models can also be derived \cite{YL:01} under
certain parametrically controlled assumptions from the \N\
developed in this section. However, the scope of these lectures
does not allow me to go into any further demonstrations.

In conclusion, I have demonstrated in detail how to derive from
scratch  the \N\ for non-interacting electrons in disordered
media, have shown some example of its perturbative usage, and in a
much more succinct mode, have illustrated how to derive and use it
in the presence of the pairing and Coulomb interactions between
electrons.

{\bf Acknowledgments.} This work has been supported by the
Leverhulme  Trust under the contract F/94/BY and by the EPSRC
grant GR/R33311.


\begin{thebibliography}{10}
\bibitem{note}
The NL$\sigma$M has been also
generalised
\cite{KhM,EF1,MirlinReview,A3S,ASO}
for ballistic systems with $L\lesssim\ell$ but such a
generalisation will not be considered in the present lectures.
\end{thebibliography}

\end{document}